\documentclass[12pt,dvipsnames]{article}
\usepackage[top=80pt,bottom=85pt,left=85pt,right=85pt]{geometry}

\pdfoutput=1
\usepackage[utf8]{inputenc}
\usepackage[T1]{fontenc} 
\usepackage[english]{babel} 
\usepackage{braket}
\usepackage[dvipsnames,table]{xcolor} %Package for fancy colors
\usepackage{physics}
\usepackage{tabularx} 
\usepackage{ragged2e}
\usepackage{rotating}
\usepackage{makecell}
\usepackage{multirow} 
\usepackage{comment} 
\usepackage{amsmath}
\usepackage{bbm}
\usepackage{cite}
\usepackage[all]{xy}

\usepackage{tikz}
\usetikzlibrary{math} 
\usetikzlibrary{fit,backgrounds}
\pgfdeclarelayer{bg}
\pgfsetlayers{bg,main}
\usetikzlibrary{3d} 
\usepackage{tikz-3dplot}
\usepackage{tikz-cd} 
\usepackage{microtype}
\usetikzlibrary{arrows.meta,positioning} 
\usepackage{amsthm}
\usepackage{mathrsfs} 
\usepackage[strict]{changepage}
\newcommand\scalemath[2]{\scalebox{#1}{\mbox{\ensuremath{\displaystyle #2}}}}

\tikzcdset{
  cells={font=\everymath\expandafter{\the\everymath\displaystyle}},
}

\usepackage{amssymb} 
\usepackage{subcaption} 
\DeclareCaptionFormat{custom}

\usepackage[pdftex,colorlinks=true]{hyperref} 
\hypersetup{urlcolor=MidnightBlue, citecolor=red, linkcolor=MidnightBlue}

\usepackage{float} 
\usepackage[textsize=tiny]{todonotes}

\usepackage[capitalise]{cleveref}

\numberwithin{equation}{section}

\definecolor{cambridgeblue}{rgb}{0.64, 0.76, 0.68}
\definecolor{caribbeangreen}{rgb}{0.0, 0.8, 0.6}
\definecolor{celadon}{rgb}{0.67, 0.88, 0.69}
\definecolor{champagne}{rgb}{0.97, 0.91, 0.81}
\definecolor{cream}{rgb}{1.0, 0.99, 0.82}
\definecolor{cyan(process)}{rgb}{0.0, 0.72, 0.92}
\definecolor{brilliantlavender}{rgb}{0.96, 0.73, 1.0}
\definecolor{candypink}{rgb}{0.89, 0.44, 0.48}

\usepackage{pdflscape}
\usepackage{paralist}

\usetikzlibrary{positioning}
\usetikzlibrary{arrows}
\usetikzlibrary{decorations.pathreplacing}
\usetikzlibrary{shapes.geometric}
\usetikzlibrary{calc}
\usetikzlibrary{decorations.pathmorphing}
\usetikzlibrary{shapes.misc}

\tikzset{gaugeSU/.style={inner sep=1.7mm,draw=none,fill=yellow,minimum size=2mm,circle, draw}}
\tikzset{flavourSU/.style={draw=none,minimum size=2mm,fill=white, regular polygon,regular polygon sides=4,draw}}
\tikzset{flavour/.style={draw=none,minimum size=0.3mm,fill=white, regular polygon,regular polygon sides=4,draw}}
\tikzset{gaugeBig/.style={inner sep=1mm,draw=none,fill=white,minimum size=2mm,circle, draw}}
\tikzset{bd/.style={circle, draw=black, inner sep=0pt, fill=black, minimum size=2mm}}
\tikzset{wd/.style={circle, draw=black, inner sep=0pt, fill=white, minimum size=2mm}}
\tikzset{Dynkin/.style={circle, draw=black, inner sep=0pt, fill=white, minimum size=2mm}}
\tikzstyle{ligne}=[draw, very thick] 
\tikzstyle{gridline}=[draw, gray] 
\tikzset{gauge/.style={circle, draw,inner sep=2.5pt,fill=white}}
\tikzset{gaugeo/.style={circle, draw,inner sep=2.5pt,fill=orange}}
\tikzset{gaugec/.style={circle, draw,inner sep=2.5pt,fill=cyan}}
\tikzset{gauger/.style={circle, draw,inner sep=2.5pt,fill=red}}
\tikzset{gaugeb/.style={circle, draw,inner sep=2.5pt,fill=blue}}
\tikzset{gaugeg/.style={circle, draw,inner sep=2.5pt,fill=green}}
\tikzset{gaugem/.style={circle, draw,inner sep=2.5pt,fill=magenta}}
\tikzset{gaugey/.style={circle, draw,inner sep=2.5pt,fill=yellow}}
\tikzset{hasse/.style={circle, fill,inner sep=2pt}}
\tikzset{shrinky/.style={circle, fill,inner sep=1pt}}
\tikzset{sized/.style={circle, draw, inner sep=1.5pt}}
\tikzset{seven/.style={circle, draw,inner sep=3pt}}

\tikzset{dotto/.style={circle, orange, draw,inner sep=1.5pt,fill=orange}}
\tikzset{dottp/.style={circle, purple, draw,inner sep=1.5pt,fill=purple}}
\tikzset{dottc/.style={circle, cyan, draw,inner sep=1.5pt,fill=cyan}}
\tikzset{dottr/.style={circle, red, draw,inner sep=1.5pt,fill=red}}
\tikzset{dottb/.style={circle, blue, draw,inner sep=1.5pt,fill=blue}}
\tikzset{dottg/.style={circle, green, draw,inner sep=1.5pt,fill=green}}
\tikzset{dottm/.style={circle, magenta, draw,inner sep=1.5pt,fill=magenta}}
\usepackage{ifthen}
\usepackage{xstring}

\usepackage{cancel}
\newcommand{\xdownarrow}[1]{%
  {\left\downarrow\vbox to #1{}\right.\kern-\nulldelimiterspace}
}
  
\usepackage{cancel}
\begin{document}

\begin{titlepage}

\phantom{wowiezowie}

\vspace{-1cm}

\begin{center}

{\Huge {\bf 5d SCFT fixtures\\
\vspace{0.2cm}}}

% \bigskip

\vspace{1cm}
{\Large Michele Del Zotto$^{\dagger\ddagger\star}$} and \Large Andrea Sangiovanni$^{\dagger\ddagger\star}$\\ 

\vspace{1cm}

{\it
{\fontsize{9pt}{10pt}\selectfont
\renewcommand{\arraystretch}{2.5}

\begin{tabular}{c}
\makecell{$^\dagger$ \text{Mathematics Institute, Uppsala University,} \\ \text{Box 480, SE-75106 Uppsala, Sweden}}\\
\makecell{$^\star$ \text{Centre for Geometry and Physics, Uppsala University,} \\ \text{Box 480, SE-75106 Uppsala, Sweden}\\} \\
\makecell{$^\ddagger$ \text{Department of Physics and Astronomy, Uppsala University,}\\ \text{Box 516, SE-75120 Uppsala, Sweden}\\}\\
\end{tabular}
}}

\vskip .5cm
{\footnotesize \tt   michele.delzotto@math.uu.se \hspace{1cm} andrea.sangiovanni@math.uu.se} \\

\vskip 1cm
     	{\bf Abstract }
\vskip .1in

\end{center}

\noindent 

In the context of the atomic classification of 5d SCFTs building blocks with non-simply laced global symmetries are lacking. In this work we fill this gap by introducing a new class of 5d UV fixed points (the 5d SCFT "fixtures"). Physically, 5d SCFTs fixtures are the result of a Higgs branch RG flow, realized geometrically by M-theory on singular threefolds which we identify explicitly. In particular, this enables the construction of 5d conformal matter SCFTs with non-simply laced flavor symmetry. Along the way, we discuss their circle reduction in relation to 4d theories of class-$\mathcal S$ type. 

%that constitute the 5-dimensional generalization of 4d $\mathcal{N}=2$ class-$\mathcal{S}$ fixtures: they are encoded by a geometric choice  that coincides with the puncture data of a fixture in class-$\mathcal{S}$, when such description exists, and they can be gauged together to produce novel 5d SCFTs. Crucially, not all of them descend to a class-$\mathcal{S}$ fixture upon circle reduction.
% Geometrically, they are engineered by  which are complex structure deformations of 5d conformal matter threefolds. Two main implications arise from this class of theories: the explicit construction of bifundamental 5d SCFTs with non-simply laced flavor symmetry, and a possible novel framework for the classification of 5d SCFTs with geometric origin.
 
\noindent

\eject

\end{titlepage}

% titlepage 
\tableofcontents

\section{Introduction}\label{sec: intro}
In the quest for a classification of higher-dimensional superconformal field theories (SCFTs) with 8 supercharges, 5d SCFTs have long posed a challenge, resisting a completely satisfactory treatment. Various approaches have enabled great progress, scanning large swaths of the 5d SCFT landscape. Results in this direction have been obtained from circle reduction from 6d $\mathcal{N}=(1,0)$ SCFTs (plus possible holonomies) as well as bottom-up constructions of ``shrinkable'' local collections of surfaces \cite{DelZotto:2017pti,Apruzzi:2019kgb,Apruzzi:2019opn,Apruzzi_2020,Jefferson_2018,Bhardwaj:2018yhy,Bhardwaj:2018vuu,Bhardwaj:2019xeg,Bhardwaj:2020gyu,Bhardwaj:2019jtr}, 5-brane web setups, possibly involving additional decorations such as orientifold planes, as well as webs related to generalized toric polygons \cite{Aharony:1997ju,Aharony:1997bh,DeWolfe:1999hj,Benini:2009gi,Bergman:2013aca,Zafrir:2014ywa,Hayashi:2015zka,Hayashi:2015fsa,Bergman:2015dpa,Hayashi:2018lyv,Hayashi:2018bkd,Hayashi:2019yxj,Hayashi_2020,Bergman:2020myx,Bourget:2023wlb,Alexeev:2024bko,Arias-Tamargo:2024fjt,CarrenoBolla:2024fxy,Collinucci:2026kom}, geometric engineering via M-theory on canonical singularities  \cite{Xie_2017,Tian:2021cif,Closset:2020scj,
Closset:2020afy, Closset:2021lwy, Collinucci:2021ofd,Mu:2023uws, Collinucci:2025rrh, Dramburg:2025tlb}, and atomic classification \cite{DeMarco:2023irn,DeMarco:2025ugw,Bourget:2026ono,DeMarco:2026tnc}.
Naturally, these technologies are tightly interconnected and completely agree in the realms where they overlap, giving rise to dual perspectives on the same 5d fixed points. In particular the atomic classification of 5d SCFTs relies on geometric engineering tools, but gives a different perspective on the problem: instead of classifying all canonical threefold singularities directly, one finds a collection of minimal ones, the atoms, and gives gluing conditions for those \cite{Heckman:2015bfa}.\\

So far in the 5d atomic classification program 5d conformal matter theories of type $\mathfrak{g}$ have been constructed for $\mathfrak{g}\in ADE$. Similarly, Trinions and Tetraon theories with global symmetries involving three or four distinct factors have been constructed for various ADE types. A few classes of examples of 5d SCFTs with global symmetries that are non-simply laced can be realized exploiting brane web constructions \cite{VanBeest:2020kxw}, 6d circle reductions \cite{Bhardwaj:2019fzv} and engineering on orbifolds of $\mathbb{C}^3$ \cite{Tian:2021cif}. It is therefore natural to ask whether 5d SCFTs with non-simply laced global symmetries arise from 5d atoms and 5d hybrids and how. Addressing this question is the main result in this paper. We present an approach based on the geometrization of 5d Higgs branch RG flows that allows to enrich significantly the landscape of building blocks, allowing also the realization of theories with \textit{non-simply laced} (including exceptional) 5d global symmetries.\\ %The classification of such descendents is an incredibly rich problem. In this paper, we content ourselves in exhibiting some examples of these constructions, leaving a systematic analysis for future work.\\

Concretely, we take as our starting point the 5d conformal matter (5d CM) theories constructed in \cite{DeMarco:2023irn}, and later expanded upon in \cite{DeMarco:2025ugw,Bourget:2026ono}. These are 5d SCFTs of arbitrarily high rank, and with flavor symmetry containing a $\mathfrak{g}\oplus\mathfrak{g}$ subalgebra, with $\mathfrak{g} \in ADE$. In this work, we show that, considering flows along the Higgs branch of 5d CM theories, 5d SCFTs with flavor symmetry containing at least a $\mathfrak{g}'\oplus\mathfrak{g}''$ subalgebra can be reached. We call these theories \textit{5d fixtures}, since they behave similarly to their renowned 4d counterparts \cite{Chacaltana:2010ks,Chacaltana:2011ze,Chacaltana:2012zy,Chacaltana:2012ch,Chacaltana:2013oka,Chacaltana:2014jba,Chacaltana:2015bna,Chacaltana:2016shw,Chacaltana:2017boe,Chacaltana:2018vhp}, as we will show. Crucially, $\mathfrak{g}'$ and $\mathfrak{g}''$ can, by a suitable choice of RG flow, be picked among \textit{any simple Lie algebra}\footnote{Not all combinations of simple Lie algebras can appear, but each algebra in the ABCDEFG classification can be in one of the summands of $\mathfrak{g}'\oplus\mathfrak{g}''$.}. All bifundamental conformal matter 5d SCFTs with flavor symmetry at least $\mathfrak{g}\oplus\mathfrak{g}$ with $\mathfrak{g}\in ABCDEFG$ can be engineered in this fashion. By translating this statement in terms of explicit Calabi-Yau threefold geometries, key features of the corresponding 5d SCFTs can be extracted, allowing an explicit check of the field theoretic statements. Since this method can be applied starting from 5d CM theories of any rank, it is manifest that a large number of 5d SCFTs, with vast freedom over their ranks and flavor symmetries, can be engineered in such fashion. In this note we wish to sketch out the salient traits of this novel approach, leaving the exploration of its full potential (and its possible repercussions on advocating for a full classification of 5d SCFTs) for future work.\\

\noindent Our main results are as follows:
\begin{itemize}
    \item We explicitly show how to construct dynamical deformations for the Calabi-Yau threefolds engineering 5d CM theories, building on the work of \cite{Collinucci:2020jqd} and \cite{DeMarco:2026jyj}. Field-theoretically, these correspond to Higgs branch RG flows. Our ansatz allows the construction of novel canonical threefolds that engineer bifundamental 5d SCFTs with $\mathfrak{g}'\oplus\mathfrak{g}''$ flavor symmetry, where $\mathfrak{g}'$ and $\mathfrak{g}''$ are simple Lie algebras. Mathematically, this amounts to the identification of a distinguished set of deformations for classes of canonical threefolds with non-isolated singularities. These are notoriously challenging to tackle, since their Milnor ring is infinite-dimensional. We expect that our proposal should come with a rigorous mathematical counterpart in deformation theory.
    \item We check the soundness of our technology by examining a few examples in full details: we take particular care in proving the presence of non-simply laced flavor symmetries in the 5d SCFT fixture by explicitly identifying flavor curves and computing their intersections with non-compact divisors, given a choice of complete resolution of the starting singular threefold.
    \item We show that a subset of 5d SCFT fixtures admit a circle reduction that is a 4d $\mathcal{N}=2$ theory engineered as a class-$\mathcal{S}$ fixture with three regular punctures, showing perfect agreement between the rank of their Coulomb branches, as well as for their flavor symmetries and rank of the Higgs branch. In this context, the flavor symmetries (including the non-simply laced ones) are predicted by the centralizers of the nilpotent orbits characterizing the class-$\mathcal{S}$ setup: as is well known, centralizers of nilpotent orbits in simply laced algebras can be non-simply laced. Not all 5d SCFT fixtures admit a class-$\mathcal{S}$ circle reduction, and we spell out such cases explicitly: these should give rise to other (as of yet unidentified) 4d $\mathcal{N}=2$ theories. We propose a criterion to identify the cases that admit a class-$\mathcal{S}$ reduction, exhibiting the explicit deformation parameters that encode the canonical threefold, thanks to the theory of nilpotent orbits. Vice versa, this puts forward a candidate 5d parent for each 4d fixture with regular punctures (subject to a condition on one of the three punctures).
    \item Analogously to their 4d cousins, 5d fixtures can be fused together, under suitable conditions. This operation, applied in the 5d CM context in \cite{DeMarco:2023irn} and \cite{Bourget:2026ono}, here gives rise to a plethora of novel theories, including further 5d SCFTs with non-simply laced flavor symmetry. For 5d fixtures with a class-$\mathcal{S}$ descendant, fusion has an obvious counterpart in terms of gluing of regular punctures.
\end{itemize}

\subsection{Outline}
We start in Section \ref{sec: 5d CM} by reviewing the construction, via M-theory geometric engineering on non-compact Calabi-Yau threefolds, of 5d bifundamental conformal matter SCFTs. We highlight their properties under gauging and Higgsing, as well as their relation with class-$\mathcal{S}$ theories. In Section \ref{sec: 5d fixtures} we lay down the core result of this work: by studying dynamical deformations of 5d conformal matter theories, we introduce 5d SCFT fixtures. The relation between 4d and 5d fixtures is shown in Section \ref{sec: fixtures and class S}. We exemplify these novel theories in Section \ref{sec: examples}, showing explicit 5d SCFT fixtures with non-simply laced flavor symmetry. We also emphasize that some 5d SCFT fixtures descend to 4d fixtures. We study the gauging of 5d fixtures in Section \ref{sec: gauging}, producing further new 5d SCFTs with non-simply laced flavor symmetry, and preliminarily explore the Higgs branch of 5d fixtures in Section \ref{sec: higgs branch}. We outline a roadmap for future work in Section \ref{sec: outlook}. The concrete resolution techniques employed in the main text are reviewed in Appendix \ref{app: resolution}, and we include additional examples of 5d fixtures in Appendix \ref{app: E8 example}, one of which displays non-simply laced exceptional flavor symmetry. In Appendix \ref{app: cDV} we review key facts about cDV threefolds, needed to systematically construct those 5d fixtures that descend to a 4d fixture.

\section{Review: 5d conformal matter SCFTs}\label{sec: 5d CM}
\indent We start by briefly reviewing 5d conformal matter (CM) SCFTs and the related threefolds, introduced in \cite{DeMarco:2023irn} and \cite{DeMarco:2025ugw}, and later expanded upon in \cite{Bourget:2026ono} and \cite{DeMarco:2026tnc}. This will grant us the leverage to show, in later sections, that 5d fixtures are engineered by CY3 that arise from \textit{dynamical complex structure deformations} of the ones employed to yield 5d CM theories.

\subsection{5d conformal matter threefolds}\label{sec: CM threefolds}
5d bifundamental conformal matter theories are 5d SCFTs whose flavor symmetry contains \textit{at least} a $\mathfrak{g}\oplus\mathfrak{g}$ subalgebra, with $\mathfrak{g}\in ADE$. This definition parallels that of 6d conformal matter SCFTs. A wide class of 5d CM theories can be engineered via M-theory on canonical threefolds of the form:
\begin{equation}\label{5d CM CY3}
    \begin{cases}
        P_{\mathfrak{g}}(x,y,z) = 0,\\
        uv=w(x,y,z),\\
    \end{cases} \quad \subset \mathbb{C}^5.
\end{equation}
where $w(x,y,z)$ is an arbitrary polynomial in $x,y,z$, and $P_{\mathfrak{g}}(x,y,z)$ is the defining polynomial of a ADE/Kleinian/Du Val singularity of type $\mathfrak{g}\in ADE$. Our conventions for the $P_{\mathfrak{g}}(x,y,z)$ are as follows:
\begin{equation}\label{ADE sing}
 \begin{cases}
P_{A_k}(x,y,z) = xy+z^{k+1}, \\
P_{D_k}(x,y,z) = x^2 +zy^2-z^{k-1}, \\
P_{E_6}(x,y,z) =x^2+y^3+\frac{z^4}{4}, \\
P_{E_7}(x,y,z) = x^2+y^3+yz^3, \\
P_{E_8}(x,y,z) = x^2+y^3+z^5,
\end{cases}
\end{equation}
It is straightforward to notice that \eqref{5d CM CY3} supports singularities of type $\mathfrak{g}$ along two non-compact complex lines intersecting at a single point, which translate into a $\mathfrak{g}\oplus\mathfrak{g}$ flavor symmetry in the corresponding 5d SCFT. The singular lines are:
\begin{equation}\label{singular lines}
    x=y=z=u=0, \quad\quad x=y=z=v=0.
\end{equation}
In general, depending on the choice of $w(x,y,z)$, there can be additional singular non-compact lines of type $A$.\\
\indent Other than by direct inspection of the singular geometry, the existence of a $\mathfrak{g}\oplus\mathfrak{g}$ flavor symmetry at the UV fixed point can be gleaned from an IR perspective: the threefolds \eqref{5d CM CY3} admit a partially resolved phase that encodes a low-energy quiver phase $\mathsf{Q}_{\mathfrak{g}}$, where the gauge nodes are special unitary and arranged as the $\mathfrak{g}$ Dynkin diagram. The arguments of \cite{Tachikawa,Yonekura:2015ksa} then prove that such IR phase flows in the UV to a 5d SCFT with $\mathfrak{g}\oplus\mathfrak{g}$ flavor symmetry. Such UV completion is precisely provided by the SCFTs engineered by \eqref{5d CM CY3}. Further details about the partially resolved phase can be found in \cite{DeMarco:2023irn}.

\subsection{5d conformal matter, gauging and Higgsing}\label{sec: review higgsing}
\indent 5d CM theories are organized into three classes:
\begin{itemize}
    \item 5d CM atoms, encoded by a CY3 \eqref{5d CM CY3} where $w$ has at least a linear term, and it is not factorizable\footnote{We consider factorizability into irreducible components over the ring $\mathbb{C}[x,y,z]$.}.
    \item 5d CM hybrids, encoded by a CY3 \eqref{5d CM CY3} where $w$ contains no linear term, and it is not factorizable.
    \item 5d CM molecules, encoded by a CY3 \eqref{5d CM CY3} where $w$ is a product of at least two factors.
\end{itemize}
This structure possesses a beautiful counterpart in terms of Lie-theoretic data, encoded by dominant coweights of the Lie algebra $\mathfrak{g}$. We refer the reader to \cite{Bourget:2026ono} for the full details.\\

\indent Physically, atoms, molecules and hybrids are related via Higgsing and gauging. In order to stay within the class of 5d CM theories (namely, to preserve the $\mathfrak{g}\oplus\mathfrak{g}$ flavor symmetry), Higgsing should correspond to a dynamical complex structure deformation of the threefolds \eqref{5d CM CY3} that does \textit{not} break the singular lines \eqref{singular lines}. Of course, this is only a small subset of all possible Higgsings: all the remaining ones are crucial to the introduction of 5d fixtures, that we initiate in Section \ref{sec: 5d fixtures}. For the time being, we review Higgsings that preserve the bifundamental $\mathfrak{g}\oplus\mathfrak{g}$ symmetry. These are realized via the following deformations:
\begin{equation}\label{5d CM deformations}
    \begin{cases}
        P_{\mathfrak{g}}(x,y,z) = 0,\\
        uv=w(x,y,z)+\mathcal{D}(x,y,z),\\
    \end{cases} \quad \subset \mathbb{C}^5,
\end{equation}
where $\mathcal{D}(x,y,z)$ is a polynomial with $\nu$ independent coefficients (with $\nu$ a finite positive integer). The $\nu$ independent deformations are related to the low-energy quiver phase $\mathsf{Q}_{\mathfrak{g}}$ recalled in Section \ref{sec: CM threefolds}:
\begin{equation}
    \nu = n_H-n_V -\text{rank}(\mathfrak{g}),
\end{equation}
where $n_H$ and $n_V$ are respectively the number of hypermultiplets and the number of vectormultiplets, that can be readily computed from the explicit presentation of the quiver. The work of \cite{Collinucci:2020jqd} has shown how to compute $\mathcal{D}(x,y,z)$: relying on the $\mathbb{C}^*$-action $(u,v)\rightarrow (\lambda u,\lambda^{-1}v)$, one can reduce \eqref{5d CM CY3} along a circle, and consider the corresponding Type IIA setup. In this setting, D6-branes are located on $w(x,y,z)=0$. \textit{Dynamical deformations} are the ones that do not perturb stacks of non-compact D6-branes (as such deformations would otherwise require an infinite amount of energy). Thanks to this ansatz, one can systematically compute $\mathcal{D}(x,y,z)$ employing tools in commutative algebra, as reviewed in \cite{Bourget:2026ono}.\\

\indent On the other hand, gauging corresponds to the gluing of two 5d CM threefolds along a $\mathbb{P}^1$: shrinking such $\mathbb{P}^1$ to zero volume yields a 5d \textit{molecule} SCFT.\footnote{\ Here we refer to this process as ``gauging'', and more properly this is a fusion of two 5d SCFTs. Fusion is an operation on higher dimensional SCFTs which generalizes to higher dimensions the notion of conformal diagonal gauging of two identical flavor symmetries in 4d (see e.g. \cite{Heckman:2018pqx,Hayashi:2019fsa}).} The $\mathbb{P}^1$ in question is obtained as the compactification of two non-compact singular lines of type \eqref{singular lines}, one for each of the two initial 5d CM threefolds. Physically, the gluing amounts to gauging a diagonal $\mathfrak{g}$ flavor symmetry between the two starting 5d CM SCFTs. Suppose that the starting threefolds are:
\begin{equation}
    \begin{cases}
        P_{\mathfrak{g}}(x,y,z)=0,\\
        uv=w_1(x,y,z),\\
    \end{cases} \quad\quad
        \begin{cases}
        P_{\mathfrak{g}}(x,y,z)=0,\\
        uv=w_2(x,y,z).\\
    \end{cases} 
\end{equation}
Then the singular equation for the 5d molecule SCFT can be read off as follows:
\begin{equation}\label{5d molecule}
        \begin{cases}
        P_{\mathfrak{g}}(x,y,z)=0,\\
        uv=w_1(x,y,z)w_2(x,y,z).\\
    \end{cases} 
\end{equation}
 Crucially, this operation can be repeated at will, producing molecules with an arbitrary high number of factors on the right-hand side of ``$uv=\ldots$''.\\
 
\indent The properties of atoms, hybrids and molecules are transparent, once one turns on a deformations in $\mathcal{D}(x,y,z)$:
\begin{itemize}
    \item atoms can only be Higgsed to other atoms;
    \item hybrids admit \textit{at least} one Higgsing to a molecule;
    \item molecules are obtained via gauging of atoms and hybrids, and can be Higgsed to other 5d CM SCFTs.
\end{itemize}
A detailed illustration of these features, along with plenty of examples, are presented in \cite{Bourget:2026ono}.

\subsection{5d conformal matter and Class-$\mathcal{S}$ theories}
It has been shown in \cite{DeMarco:2025ugw} and \cite{Bourget:2026ono} that 5d CM atoms admit a circle reduction that descends to a 4d $\mathcal{N}=2$ SCFT which is a class-$\mathcal{S}$ fixture, i.e.\ a class-$\mathcal{S}$ setup on a sphere with three regular punctures. We denote the triple of regular punctures defining an atom of type $\mathfrak{g}$ as:
\begin{equation}
(\mathcal{O}_{\text{max}},\mathcal{O}_{\text{max}},\mathcal{O}_{\text{rest}}),
\end{equation}
where we employ the standard notation for nilpotent orbits. $\mathcal{O}_{\text{max}}$ denotes the full puncture, corresponding to the maximal nilpotent orbit of $\mathfrak{g}$ in Hitchin notation, and $\mathcal{O}_{\text{rest}}$ denotes the orbit encoding the third puncture, which is specified by $w(x,y,z)$ in \eqref{5d CM CY3}.\\
\indent Hybrids and molecules do not admit a circle reduction described by a class-$\mathcal{S}$ setup with regular punctures. However, molecules admit a 5d low-energy phase that can be reduced to class-$\mathcal{S}$ with two full punctures, along with a bunch of other regular punctures\footnote{Each minimal puncture is related to the factors $w_i(x,y,z)$ in the gauging \eqref{5d molecule}.}.\\
\indent The two $\mathcal{O}_{\text{max}}$ punctures encode the $\mathfrak{g}\oplus\mathfrak{g}$ symmetry of the resulting 4d $\mathcal{N}=2$ SCFT, inherited from its 5d parent. $\mathcal{O}_{\text{rest}}$ dictates extra symmetry factors $\mathfrak{g}_{\text{rest}}$. The case $\mathfrak{g}=A$ is the only one where there exists a choice of $w(x,y,z)$ such that $\mathcal{O}_{\text{rest}}=\mathcal{O}_{\text{max}}$. For $\mathfrak{g}=D,E$, one obtains at most a proper subalgebra of $\mathfrak{g}$ from $\mathcal{O}_{\text{rest}}$. The allowed choices of $\mathcal{O}_{\text{rest}}$, namely the choices that correspond to a singular threefold that engineers the UV fixed point, are listed in Appendix C of \cite{Bourget:2026ono}. The deformations appearing in \eqref{5d CM deformations} amount in 4d to a partial closure of the puncture $\mathcal{O}_{\text{rest}}$, since they leave the $\mathfrak{g}\oplus\mathfrak{g}$ symmetry untouched. As we have mentioned, this is clearly only a partial exploration of the 5d UV Higgs branch. It can be argued employing the tools of \cite{Chacaltana:2012zy} that the Higgs branch dimension of the 4d class-$\mathcal{S}$ theories at hand is:
\begin{equation}
    \text{dim}_{\mathbb{H}}^{\text{class}-\mathcal{S}}(HB) = \nu + \text{dim}_{\mathbb{C}}(\mathfrak{g}).
\end{equation}
Since the dimension of the Higgs branch is preserved under dimensional reduction, this implies:
\begin{equation}
   \text{dim}_{\mathbb{H}}^{5d \text{ SCFT}}(HB) =  \text{dim}_{\mathbb{H}}^{\text{class}-\mathcal{S}}(HB). 
\end{equation}
Hence the deformed threefold \eqref{5d CM deformations} is not capturing the full UV Higgs branch: $\text{dim}_{\mathbb{C}}(\mathfrak{g})$ dynamical deformations are not accounted for. An analogous observation is valid even for hybrids and molecules, despite the lack of a class-$\mathcal{S}$ description. In the next Section we show how to identify the missing deformations, with far-reaching consequences.

\section{5d fixtures}\label{sec: 5d fixtures}
In this Section we introduce a class of non-compact canonical threefolds that, through M-theory geometric engineering, act as \textit{5d fixtures}, i.e.\ a set of 5d SCFTs labelled by a choice of deformation data that encodes the engineering threefold, that can be gauged together under suitable constraints, as outlined in the Introduction. \\
As recalled in the previous Section, the 5d SCFTs engineered by the Calabi-Yau threefolds \eqref{5d CM CY3} sport a $\nu+\text{dim}_{\mathbb{C}}(\mathfrak{g})$ dimensional Higgs branch. Correspondingly, the Calabi-Yau threefolds \eqref{5d CM CY3} must admit $\nu+\text{dim}_{\mathbb{C}}(\mathfrak{g})$ dynamical deformations, where $\nu$ depends on the specific choice of $w(x,y,z)$ and, in the case of atoms, is related to the puncture $\mathcal{O}_{\text{rest}}$. The fully deformed threefold should hence look like:
\begin{equation}\label{5d CM full deformations}
    \begin{cases}
        P_{\mathfrak{g}}(x,y,z) + \mathcal{F}(u,v,y,z) = 0,\\
        uv=w(x,y,z)+\mathcal{D}(x,y,z),\\
    \end{cases} \quad \subset \mathbb{C}^5,
\end{equation}
where $\mathcal{F}(u,v,y,z)$ does not depend on $x$, since it always appears quadratically (or is involved in quadratic terms) in $P_{\mathfrak{g}}(x,y,z)$. There are $\nu$ independent deformation coefficients in $\mathcal{D}(x,y,z)$, and $\text{dim}_{\mathbb{C}}(\mathfrak{g})$ independent deformation coefficients in $\mathcal{F}(u,v,y,z)$, thus accounting for the full UV Higgs branch dimension of the starting 5d SCFT. Identifying the dynamical deformations $\mathcal{F}(u,v,y,z)$ is a non-trivial task, since it amounts to deformations of non-isolated singular lines in a canonical threefold (analogously, to the partial closure of the two full punctures in the class-$\mathcal{S}$ setup for atoms), for which the Milnor number is not well-defined. We now outline how to identify $\mathcal{F}(u,v,y,z)$ systematically, building on Lie-theoretic and physics arguments.\\

\indent Recall that the versal deformations of Du Val singularities can be written as \cite{Katz:1992aa}:
\begin{equation}\label{Du Val versal}
\begin{cases}
V_{A_n}(x,y,z) = xy+z^{n+1}+\sum_{i=2}^{n+1}\alpha_i z^{n+1-i}, \\
V_{D_n}(x,y,z) = x^2 +zy^2-z^{n-1}-\sum_{i=1}^{n-1}\delta_{2i}z^{n-i-1}+2\gamma_n y, \\
V_{E_6}(x,y,z) =x^2+y^3+\frac{z^4}{4}+\epsilon_2 y z^2+\epsilon_5 yz+\epsilon_6 z^2+\epsilon_8 y+\epsilon_9 z+\epsilon_{12}, \\
V_{E_7}(x,y,z) = x^2+y^3+yz^3+\tilde{\epsilon}_2y^2z+\tilde{\epsilon}_6y^2+\tilde{\epsilon}_8 yz+\tilde{\epsilon}_{10}z^2+\tilde{\epsilon}_{12}y+\tilde{\epsilon}_{14}z+\tilde{\epsilon}_{18}, \\
V_{E_8}(x,y,z) = x^2+y^3+z^5+\hat{\epsilon}_{2}yz^3+\hat{\epsilon}_{8}yz^2+\hat{\epsilon}_{12}z^3+\hat{\epsilon}_{14}yz+\hat{\epsilon}_{18}z^2+\hat{\epsilon}_{20}y+\hat{\epsilon}_{24}z+\hat{\epsilon}_{30},
\end{cases}
\end{equation}
where the $\alpha_i,\delta_i,\gamma_n,\epsilon_i,\tilde{\epsilon}_i$ and $\hat{\epsilon}_i$ are the deformation parameters. The base space of the versal deformation is isomorphic to $\mathfrak{t}/\mathcal{W}$, where $\mathfrak{t}$ is the Cartan subalgebra of $\mathfrak{g}$, and $\mathcal{W}$ is its Weyl group. The subscripts of the versal deformation parameters then express their degree, interpreted as polynomials depending on variables in $\mathfrak{t}$. Relatedly, notice that the defining polynomial of a Du Val singularity is quasi-homogeneous under the $\mathbb{C}^*$-action:
\begin{equation}\label{C-star weights}
    P_{\mathfrak{g}}(\lambda^{w_x}x,\lambda^{w_y}y,\lambda^{w_z}z)\longrightarrow \lambda^{h_{\mathfrak{g}}} P_{\mathfrak{g}}(x,y,z),
\end{equation}
with $h_{\mathfrak{g}}$ the Coxeter number of $\mathfrak{g}$. Then the subscripts of $\alpha_i,\delta_i,\gamma_n,\epsilon_i,\tilde{\epsilon}_i$  are the degrees of the deformation parameters needed to preserve the quasi-homogeneity of $V_{\mathfrak{g}}(x,y,z)$ under \eqref{C-star weights}. Take $\mathfrak{g}=E_6$ as an example: given $h_{E_6}=12$, it is easy to check that:
\begin{equation}
    w_x = 6,\quad w_y = 4, \quad w_z = 3.
\end{equation}
Now consider the versal deformation term $\epsilon_2yz^2$. The monomial $yz^2$ has weight 10, and hence $\epsilon_2$ must have weight 2 under the $\mathbb{C}^*$-action, in order to preserve the quasi-homogeneity of $V_{E_6}(x,y,z)$. A similar reasoning applies to the other versal deformation terms.\\ 
\indent Define as $f_i^{\mathfrak{g}}(y,z)$ the monomials appearing in the deformation terms of $V_{\mathfrak{g}}(x,y,z)$, with $i$ the degree of the corresponding deformation coefficient. E.g.\ for $\mathfrak{g}=E_6$ we have:
\begin{equation}
    f_2^{E_6} = yz^2, \quad f_5^{E_6} = yz , \quad f_6^{E_6} = z^2 , \quad f_8^{E_6} = y , \quad f_9^{E_6} = z , \quad f_{12}^{E_6} = 1.
\end{equation}

In order to detect the deformations $\mathcal{F}(u,v,y,z)$, notice that we expect a symmetry $u\leftrightarrow v$ that exchanges the two singular lines \eqref{singular lines}, and consequently the two $\mathfrak{g}$ flavor factors. Denote as $|\Delta|$ the number of roots in the $\mathfrak{g}$ Lie algebra. We claim that the generic $\mathcal{F}(u,v,y,z)$ is a polynomial of the form:
\begin{equation}\label{F polynomial}
    \mathcal{F}(u,v,y,z) = \sum_{i_1}\sum_{j=1}^{i_1-1} c_j^{(i_1)} u^j\cdot f_{i_1}^{\mathfrak{g}}(y,z)+ \sum_{i_2}\sum_{j=1}^{i_2-1} c_j^{(i_2)} v^j\cdot f_{i_2}^{\mathfrak{g}}(y,z)+ \sum_{i_3} c^{(i_3)} \cdot f_{i_3}^{\mathfrak{g}}(y,z).
\end{equation}
Namely, $\mathcal{F}(u,v,y,z)$ has the following properties:
\begin{enumerate}\label{enumerate def}
    \item it contains $\frac{|\Delta|}{2}$ deformations of the form $u^j\cdot f_{i_1}^{\mathfrak{g}}(y,z)$, with $f_{i_1}^{\mathfrak{g}}(y,z)$ the functions defined above, and $j=1,\ldots,i_1-1$.
    \item it contains $\frac{|\Delta|}{2}$ deformations of the form $v^j\cdot f_{i_2}^{\mathfrak{g}}(y,z)$, with $f_{i_2}^{\mathfrak{g}}(y,z)$ the functions defined above, and $j=1,\ldots,i_2-1$.
    \item it contains $\text{rank}(\mathfrak{g})$ deformations of the form $f_{i_3}^{\mathfrak{g}}(y,z)$.
\end{enumerate}
In total there are $\frac{|\Delta|}{2}+\frac{|\Delta|}{2}+\text{rank}(\mathfrak{g}) = \text{dim}(\mathfrak{g})$ deformations, perfectly matching the expectation from the 5d SCFT side. Furthermore, notice that for $\mathfrak{g}=A_n$ the prescription above parallels the explicit deformation terms identified in \cite{DeMarco:2026jyj}.
Physically we obtain that:
\begin{enumerate}
    \item turning on only some of the deformations at point (1) of the list above, while turning off all the other ones, deforms only the singular line $x=y=z=v=0$ in \eqref{singular lines}. Hence the symmetry factor $\mathfrak{g}$ related to that line is Higgsed to:
    \begin{equation*}
        \mathfrak{g} \rightarrow \mathfrak{g}_1, \quad  \quad\mathfrak{g}_1 \subseteq \mathfrak{g}.
    \end{equation*}
    \item turning on only some of the deformations at point (2) of the list above, while turning off all the other ones, deforms only the singular line $x=y=z=u=0$ in \eqref{singular lines}. Hence the symmetry factor $\mathfrak{g}$ related to that line is Higgsed to:
    \begin{equation*}
        \mathfrak{g} \rightarrow \mathfrak{g}_2, \quad  \quad\mathfrak{g}_2 \subseteq \mathfrak{g}.
    \end{equation*}
    \item turning on only some of the deformations at point (3) of the list above, while turning off all the other ones, deforms both singular lines $x=y=z=u=0$ and $x=y=z=v=0$ in \eqref{singular lines} simultaneously. Hence the symmetry factors $\mathfrak{g}\oplus\mathfrak{g}$ related to those lines are Higgsed to:
    \begin{equation*}
        \mathfrak{g}\oplus\mathfrak{g} \rightarrow \mathfrak{g}_3\oplus\mathfrak{g}_3, \quad  \quad\mathfrak{g}_3 \subseteq \mathfrak{g}.
    \end{equation*}
\end{enumerate}
For 5d CM SCFTs admitting a class-$\mathcal{S}$ description, deformations at point (1) and (2) of List \ref{enumerate def} can produce a (partial) closure of the two maximal punctures (independently from each other).
%Deformations at point (3) can produce a partial closure of both maximal punctures at the same time. 
This observation brings us to the main result of this work:\\

The deformed canonical threefolds in \eqref{5d CM full deformations} engineer 5d SCFTs that we call \textit{5d fixtures}, with flavor symmetry at least $\mathfrak{g}'\oplus\mathfrak{g}''\oplus\mathfrak{g}'''$ (with $\mathfrak{g}',\mathfrak{g}''\subseteq \mathfrak{g}$, and $\mathfrak{g}'''\subseteq \mathfrak{g}_{\text{rest}}$), where $\mathfrak{g}'\oplus\mathfrak{g}''\oplus\mathfrak{g}'''$ depend on the combination of deformation parameters $\mathcal{F}(u,v,y,z)$ (namely a choice of coefficients in \eqref{F polynomial}) and $\mathcal{D}(x,y,z)$ that have been turned on. Notice that $\mathcal{D}(x,y,z)$ can always be absorbed in a re-definition of $w(x,y,z)$: the data specifying a 5d fixture is then a specific choice of the triple $(\mathfrak{g},\mathcal{F}(u,v,y,z),w(x,y,z))$.\\

The term ``5d fixtures'' is fitting in the sense that the flavor symmetry of these 5d SCFTs closely tracks the features of class-$\mathcal{S}$ fixtures. As a by-product, we will prove that 5d fixtures can produce flavor symmetry factors that are \textit{non-simply laced}. In particular, each of the ADE algebra factors associated to the non-isolated lines can give rise to the following flavor symmetries after turning on dynamical complex structure deformations:
\begin{equation}
    \begin{array}{ccc}
       A &  \quad \rightarrow \quad &  A,C\\
       D &  \quad \rightarrow \quad &  A,B,C,D \\
       E_6 &  \quad \rightarrow \quad & A,B,C,G_2\\
       E_7 &  \quad \rightarrow \quad & A,B,C,D,G_2,F_4\\
       E_8 &   \quad \rightarrow \quad & A,B,C,D,E_6,E_7,G_2,F_4\\
    \end{array}
\end{equation}
All algebras in the $ABCDEFG$ classification can then be reached from 5d fixtures, choosing a starting ADE algebra of appropriate rank.\\

We will later show that the analogy with 4d fixtures remains valid also for the gauging of 5d fixtures, which mimics the gauging of 4d fixtures. Most importantly, though, we will observe that the circle reduction of 5d fixtures \textit{comprises} 4d fixtures, but also produces examples of 4d $\mathcal{N}=2$ theories not captured by 4d fixtures. We further put forward a criterion to classify the cases that admit a class-$\mathcal{S}$ reduction, showing the explicit deformation parameters thanks to the theory of nilpotent orbits.

\section{5d fixtures and class-$\mathcal{S}$ theories}\label{sec: fixtures and class S}
In this Section we put the connection between 5d fixtures and class-$\mathcal{S}$ theories on rigorous ground. We have seen that the deformations $\mathcal{D}(x,y,z)$ in \eqref{5d CM full deformations} can be computed thanks to the methods of \cite{Collinucci:2020jqd}, corresponding to the (partial) closure of $\mathcal{O}_{\text{rest}}$, for those theories that admit a class-$\mathcal{S}$ description. Here, given a 5d CM atom encoded by a threefold \eqref{5d CM CY3}, we wish to identify which of its dynamical deformations $\mathcal{F}(u,v,y,z)$ in \eqref{5d CM full deformations} correspond to a class-$\mathcal{S}$ theory. To this end, consider a class-$\mathcal{S}$ theory corresponding to a fixture $(\mathcal{O}_{\text{max}},\mathcal{O}_{\text{max}},\mathcal{O}_{\text{rest}})$, corresponding to the circle reduction of a 5d CM atom. In the threefold geometry, the two non-compact singular lines parametrized by $u$ and $v$ correspond to the two maximal punctures. Now consider (partially) closing the two maximal punctures, flowing to the 4d theory:
\begin{equation}\label{class S flow}
(\mathcal{O}_{\text{max}},\mathcal{O}_{\text{max}},\mathcal{O}_{\text{rest}})\quad \longrightarrow \quad (\mathcal{O}_{u},\mathcal{O}_{v},\mathcal{O}_{\text{rest}})
\end{equation}
Clearly, up to symmetric swap of the variables, the closure $\mathcal{O}_{max} \longrightarrow \mathcal{O}_u$ ($\mathcal{O}_{max} \longrightarrow \mathcal{O}_{v}$) must be encoded by a complex structure deformation of the threefold depending on the variable $u$ ($v$). Notice that the 4d theory encoded by \eqref{class S flow} could be bad (or broken \cite{Comi:2025zwu}), for some choice of the orbits: from the point of view of the 5d theory this does not pose a problem, since all dynamical complex structure deformations are on the same footing\footnote{This raises the question of how to interpret the 4d descendants of 5d fixtures that would correspond to ``bad'' or ``broken'' 4d fixtures. We leave this task for future work.}.
We can then write the deformed threefold that is reached after the Higgs branch flow in \eqref{class S flow} as:
\begin{equation}\label{deformed cDV full threefold}
    \begin{cases}
        P_{\mathfrak{g}}(x,y,z) + u\mathcal{F}_u(u,y,z) + v\mathcal{F}_v(v,y,z) = 0,\\
        uv=w(x,y,z),\\
    \end{cases} \quad \subset \mathbb{C}^5.
\end{equation}

The precise prescription to relate the nilpotent orbit closure to the 5d threefold geometry can be gleaned by noticing that the threefold
\begin{equation}\label{cDV deformation}
  P_{\mathfrak{g}}(x,y,z) + u\mathcal{F}_u(u,y,z) =0 \quad (\text{respectively } P_{\mathfrak{g}}(x,y,z) + v\mathcal{F}_v(v,y,z) =0)  
\end{equation}
 is a compound Du Val (cDV) singularity, namely a one parameter deformation of the Du Val singularity $P_{\mathfrak{g}}(x,y,z)=0$. In order to determine the explicit form of \eqref{cDV deformation} we can employ the techniques developed in \cite{Collinucci:2021ofd,DeMarco:2022dgh}. Consider the orbits $\mathcal{O}^L_u$ and $\mathcal{O}^L_{v}$, which are the \textit{Spaltenstein duals} to $\mathcal{O}_u$ and $\mathcal{O}_{v}$. The logic is straightforward: thanks to the Jacobson-Morozov theorem, each of the nilpotent orbits $\mathcal{O}^L_u$ and $\mathcal{O}^L_{v}$ encodes a standard triple, defined up to conjugation. For the sake of clarity, let us focus on $\mathcal{O}^L_u$, denoting with $(X,Y,H)$ its associated standard triple. The discussion for $\mathcal{O}^L_v$ is identical. The elements of the standard triple satisfy the commutation relations:
 \begin{equation}
     [H,X]=2X, \quad [H,Y] = -2Y, \quad [X,Y]=H.
 \end{equation}
 A representative in the conjugacy class of the nilpositive element $X$ can be explicitly built for each ADE algebra, employing the standard recipe in \cite{collingwood1993}. Then, one can consider the \textit{Slodowy slice} $S_X$ through $X$, defined as:
 \begin{equation}
  S_X = \{ Z\in \mathfrak{g} \hspace{0.1cm} | \hspace{0.1cm} [Z-X,Y]=0\}.
 \end{equation}
 The Slodowy slice depends on complex parameters $c_i$, $i = 1,\ldots,\text{dim}(S_X)$.
 Define $\Phi_u$ as an auxiliary element in $\mathfrak{g}$ (called the ``Higgs field'' in \cite{Collinucci:2021ofd,DeMarco:2022dgh}, given that it acts as a vev for D6-branes in a suitable Type IIA description):
 \begin{equation}\label{higgs field}
     \Phi_u = X+ S_X \big|_{c_i \rightarrow c_i u},
 \end{equation}
 where we have chosen a linear dependence of the parameters $c_i$ on variable $u$ of the threefold. The relevant point is that the Casimirs of $\Phi_u$ encode the cDV threefold \eqref{cDV deformation}. In this way, $\mathcal{F}_u(u,y,z)$ is identified. We leave the details, that review the work of \cite{Collinucci:2021ofd,DeMarco:2022dgh}, in Appendix \ref{app: cDV}. This brings us to the key result of this section:\\

 \indent \textit{The threefold \eqref{deformed cDV full threefold}, which descends upon circle reduction to the class-$\mathcal{S}$ theory encoded by the endpoint of the flow \eqref{class S flow}, can be written down explicitly by constructing the Higgs fields $\Phi_u$ and $\Phi_v$ defined in \eqref{higgs field}, and relating their Casimirs to the threefold equation as detailed in Appendix \ref{app: cDV}.}\\

 We present an explicit example to clarify this result. Consider the 5d CM atom engineered by the threefold, which is a base-change of the $D_4$ Du Val singularity:
 \begin{equation}
     \begin{cases}
         x^2+zy^2-z^3=0,\\
         uv=x,\\
     \end{cases}
 \end{equation}
 that descends to the class-$\mathcal{S}$ theory encoded by the fixture:
 \begin{equation}
     ([7,1],[7,1],[3^2,1^2]).
 \end{equation}
Suppose that we partially close one of the punctures, reaching the fixture:
\begin{equation}\label{class S D4 example}
    ([7,1],[5,1^3],[3^2,1^2]).
\end{equation}
We can compute the Slodowy slice through a nilpositive representative of the Spaltenstein dual of $[5,1^3]$, which is $[3,1^5]$, and impose a dependence of its parameters on $u$. Explicitly, it takes the form:
\begin{equation}
    \Phi_u = \scalemath{0.85}{\left(
\begin{array}{cccc|cccc}
 0 & 1 & 0 & 0 & 0 & 1 & 0 & 0 \\
 -\frac{c_1 u}{4}  & 0 & 0 & 0 & -1 & 0 & 0 & 0 \\
 0 & 0 & 0 & 0 & 0 & 0 & 0 & 0 \\
 0 & 0 & 0 & 0 & 0 & 0 & 0 & 0 \\
 \hline
 0 & \frac{c_1 u}{4} & 0 & 0 & 0 & \frac{c_1 u}{4} & 0 & 0 \\
 -\frac{c_1 u}{4} & 0 & 0 & 0 & -1 & 0 & 0 & 0 \\
 0 & 0 & 0 & 0 & 0 & 0 & 0 & 0 \\
 0 & 0 & 0 & 0 & 0 & 0 & 0 & 0 \\
\end{array}
\right)},
\end{equation}
where we have made a specific choice of parameters, and the quadratic form of $D_4$ is:
\begin{equation}
    Q = \left(\begin{array}{cc}
       0 & \mathbbm{1}_{4\times 4}  \\
     \mathbbm{1}_{4\times 4}     & 0 
    \end{array}
    \right).
\end{equation}
Then, the Casimirs of the Slodowy slice produce, via the spectral equation \eqref{spectral equations}, the cDV threefold:
\begin{equation}
    x^2+zy^2-z^3+c_1uz^2 = 0, \quad \Rightarrow \mathcal{F}_u(u,y,z) = c_1z^2.
\end{equation}
Plugging it back into the threefold expression \eqref{deformed cDV full threefold} we finally obtain:
 \begin{equation}\label{deformed D4 example}
     \begin{cases}
         x^2+zy^2-z^3+c_1uz^2=0,\\
         uv=x.\\
     \end{cases}
 \end{equation}
The threefold \eqref{deformed D4 example} engineers a 5d SCFT fixture that descends to the class-$\mathcal{S}$ fixture \eqref{class S D4 example}, which we will check explicitly in the next Section by performing a crepant resolution. As expected from class-$\mathcal{S}$, we will see that \textit{the flavor symmetry of the 5d theory is dictated by the centralizer of the Spaltenstein dual orbits} $\mathcal{O}^L_u$ and $\mathcal{O}^L_v$. Reasoning in this fashion, the canonical threefold that produces the 5d SCFT that is the parent of a 4d fixture can be identified \textit{for each nilpotent orbit closure, and for each ADE algebra}. This of course holds only for 4d fixtures that admit a 5d origin, namely the ones with admissible $\mathcal{O}_{\text{rest}}$. As we have mentioned, the list of $\mathcal{O}_{\text{rest}}$ that can be related to a 5d theory are listed in \cite{Bourget:2026ono}. Vice versa, each canonical threefold that takes a form which can be obtained as described above, admits a class-$\mathcal{S}$ interpretation upon circle reduction.

\section{Examples of 5d fixtures}\label{sec: examples}
In this Section we display examples that showcase the crucial properties of 5d fixtures:
\begin{itemize}
    \item the appearance of non-simply laced flavor groups in M-theory geometric engineering on a canonical threefold. It can be predicted from the theory of nilpotent orbits, and it is checked via explicit crepant resolutions;
    \item the matching of the circle reduction of some 5d fixtures with class-$\mathcal{S}$ fixtures.
\end{itemize}
We address the first point by performing an explicit complete crepant resolution of the threefolds that engineer 5d fixtures. The resolution produces a set of exceptional compact divisors $S_i$, $i=1,\ldots,r$, with $r$ the rank of the 5d SCFT, as well as non-compact divisors $D_j$, $j=1,\ldots d$. The exceptional non-compact divisors give rise to non-abelian flavor symmetry factors, produced by non-compact lines of singularity, such as the ones in \eqref{singular lines}. Define the curves lying at the intersection of compact and non-compact divisors \cite{Apruzzi:2019kgb,Apruzzi:2019opn,Tian:2021cif}:
\begin{equation}\label{flavor curves}
    C_{j} = D_j \cdot(\sum_i \xi_{ij}S_i),
\end{equation}
with $\xi_{ij}$ some coefficients that depend on the choice of crepant resolution, fixed by requiring that $C_j$ supports a W-boson (namely, that its normal bundle is the unobstructed $\mathcal{O}\oplus\mathcal{O}(-2)$ bundle).
The curves in \eqref{flavor curves} are the flavor curves, since M2-branes wrapped on them engineer the W-bosons for the flavor algebra. The normal bundle of their irreducible components is $\mathcal{O}(0)\oplus\mathcal{O}(-2)$, given that there exists a flat direction extending along the non-compact exceptional divisor $D_j$. Then the symmetrized Cartan matrix of the flavor algebra is:
\begin{equation}\label{sym cartan}
    \mathcal{C}_{jk} = -D_j\cdot D_k \cdot(\sum_i \xi_{ij}S_i).
\end{equation}
The symmetrized Cartan matrix is obtained from the usual Cartan matrix by acting on it with a diagonal matrix: its diagonal entries are in general not all equal, and this signals the presence of roots of different lengths, and hence of a non-simply laced symmetry algebra.\\

\indent As regards the reduction of 5d fixtures to 4d, we follow the recipe laid down in \cite{Martone:2021drm} and concretely applied to this setting in \cite{DeMarco:2025ugw}. We claim that a 5d fixture obtained as the deformation along the Higgs branch of a 5d CM atom descends to a 4d class-$\mathcal{S}$ fixture if the following consistency checks are satisfied:
\begin{itemize}
    \item The ranks of the 5d and the 4d theories coincide. This data is extracted from the resolution of the threefold $X$, since:
    \begin{equation*}
        r = \text{dim}H_4(X,\mathbb{R}).
    \end{equation*}
    \item The flavor symmetry ranks of the 5d and 4d theories coincide. The rank of the 4d flavor symmetry is computed from the data of the punctures, while the rank $f$ of the 5d SCFT flavor symmetry is given by:
    \begin{equation*}
        f = \text{dim}H_2(X,\mathbb{R})-\text{dim}H_4(X,\mathbb{R}),
    \end{equation*}
    which can be extracted from the crepant resolution of the threefold $X$.
    \item The dimensions of the Higgs branch of the 5d and 4d theory coincide.
\end{itemize}

\subsection{5d fixture that engineers non-simply laced flavor symmetry}\label{sec: 5d fixture non simply laced}
Consider the canonical threefold in \eqref{5d CM CY3} with $\mathfrak{g}=D_4$, along with a choice of $w(x,y,z)$:
\begin{equation}\label{D4 example 1}
\begin{cases}
    x^2+zy^2-z^3=0,\\
    uv=x.
\end{cases}
\end{equation}
The 5d SCFT engineered by \eqref{D4 example 1} is a theory of rank 5, admitting a low-energy quiver phase described in \cite{DeMarco:2023irn}. This can be proven explicitly performing a crepant resolution of the singular threefold, as explained in the cited reference. It is a 5d CM atom, and upon circle reduction it descends to a class-$\mathcal{S}$ theory given by the 6d $\mathcal{N}=(2,0)$ $D_4$ SCFT on a sphere with three regular punctures, specified by the triple of nilpotent orbits:
\begin{equation}
    ([7,1],[7,1],[3^2,1^2]),
\end{equation}
where we employ the standard Hitchin notation for nilpotent orbits \cite{collingwood1993}, with powers representing repeated entries (i.e. $[3^2,1^2]=[3,3,1,1]$). Correspondingly, the flavor symmetry of the 5d SCFT is:
\begin{equation}
    D_4 \oplus D_4 \oplus \mathfrak{u}(1)\oplus\mathfrak{u}(1),
\end{equation}
barring non-abelian completions.\\
\indent We can employ the machinery reviewed in Section \ref{sec: review higgsing} and developed in Section \ref{sec: 5d fixtures} to explicitly compute the dynamical complex structure deformations of the threefold \eqref{D4 example 1}. Schematically, we obtain:
\begin{equation}\label{def par D4}
\begin{split}
 &  \mathcal{D}(x,y,z) \ni \quad(z,y,1), \\
  & \mathcal{F}(u,v,y,z) \ni \quad (z^2,uz^2,vz^2,z,uz,vz,u^2z,v^2z,u^3z,v^3z,1,u,v,u^2,v^2,\\
 & \hspace{3.2cm}  u^3,v^3, u^4,v^4,u^5,v^5,y,uy,vy,u^2y,v^2y,u^3y,v^3y), \\
\end{split}
\end{equation}
where we have indicated the deformation monomials in the brackets. The deformed threefold is then constructed taking a linear combination of the deformation monomials, with independent complex coefficients.\\

For the sake of concreteness, let's pick the following choice of deformation parameters (for generic $c_i$):
\begin{equation}\label{D4 example 1 def}
\begin{cases}
    x^2+zy^2-z^3 + c_1uz^2+c_2vz^2+c_3z^2=0,\\
    uv=x.
\end{cases}
\end{equation}
where $c_1,c_2,c_3$ are complex coefficients. Given the dictionary between deformations and non-compact singular lines above, we expect to break both $D_4$ symmetry factors. We can check it by examining the singular lines of the deformed threefold:
\begin{equation}
   L_1:\quad x=y=z=u=0, \quad\quad L_2: \quad x=y=z=v=0.
\end{equation}
The generic fiber along $L_1$ is (setting e.g. $v=1$, and substituting $x=uv$):
\begin{equation}\label{D3 sing}
   u^2+zy^2+c_1uz^2+(c_2+c_3)z^2=0. 
\end{equation}
which naively looks like a $D_3\cong A_3$ singularity. The same happens along $L_2$, provided that $u$ and $v$ are exchanged. However, it is well-known (see e.g.\ \cite{Tian:2021cif}) that the apparent Du Val fiber along a non-compact line does not necessarily correspond to the flavor symmetry engineered by that line, since the crepant resolution might produce fewer exceptional non-compact divisors than the rank of the naive flavor symmetry. Such case is the hallmark of the appearance of a non-simply laced flavor symmetry. As we have argued above, one can prove this rigorously by computing the symmetrized Cartan matrix \eqref{sym cartan}.

\begin{figure}[H]
\begin{center}
    \scalebox{0.65}{\begin{tikzpicture}
        \draw[thick] (0,0) circle (1);
        \node at (0,0) {\small$S_1$};
        \draw[thick] (0.75,0.75)--(1.75,1.75);
        \draw[thick] (2.5,2.5) circle (1);
        \node at (2.5,2.5) {\small$D_1$};
        \draw[thick] (-0.75,0.75)--(-1.75,1.75);
        \draw[thick] (-2.5,2.5) circle (1);
        \node at (-2.5,2.5) {\small$D_4$};
        \draw[thick] (0.75,-0.75)--(1.75,-1.75);
        \draw[thick] (2.5,-2.5) circle (1);
        \node at (2.5,-2.5) {\small$D_3$};
        \draw[thick] (-0.75,-0.75)--(-1.75,-1.75);
        \draw[thick] (-2.5,-2.5) circle (1);
        \node at (-2.5,-2.5) {\small$D_2$};
        \draw[thick] (2.5,1.45)--(2.5,-1.45);
        \draw[thick] (-2.5,1.45)--(-2.5,-1.45);
        \end{tikzpicture}}
        \caption{Exceptional locus of the crepant resolution of the threefold \eqref{D4 example 1 def}.}
        \label{fig: D4 example 1}
\end{center}
\end{figure}
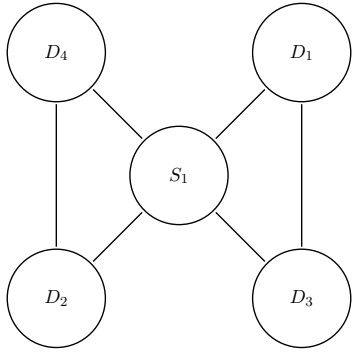

We claim that the new threefold engineers a rank 1 5d SCFT with a rank 7 flavor symmetry, containing \textit{at least} a $B_2 \oplus B_2$ subalgebra,
up to non-abelian completions. This is to be expected since the deformation terms "$c_1uz^2$" and "$c_2vz^2$" arise from nilpotent orbit closures labelled by $[5,1^3]$, as shown in Section \ref{sec: fixtures and class S}. We prove it by explicitly performing a complete crepant resolution of the singular threefold. We present the detailed description of the crepant resolution in Appendix \ref{app: resolution}. The upshot is that the resolution inflates 1 compact exceptional divisor $S_1$, along with 4 non-compact exceptional divisors $D_j$, $j=1,2,3,4$. The intersection pattern between the divisors is depicted in Figure \ref{fig: D4 example 1}.

The divisors $D_1,D_3$ arise from the resolution of $L_1$, while $D_2,D_4$ arise from the resolution of $L_2$. Notice that there are less non-compact divisors than the ones expected from the naive ADE fiber in \eqref{D3 sing}.
Consider the flavor curves:
\begin{equation}\label{D4 flavor curves 1}
    C_1 = D_1\cdot S_1, \quad C_3 = D_3\cdot S_1, \quad C_2 = D_2\cdot S_1, \quad C_4 = D_4\cdot S_1.
\end{equation}
The symmetrized Cartan matrices for the flavor symmetries supported on $L_1$ and $L_2$ are then:
\begin{equation}
    \mathcal{C}_1 = -\begin{pmatrix}
        C_1\cdot D_1 & C_1 \cdot D_3 \\
        C_3\cdot D_1 & C_3 \cdot D_3 \\
    \end{pmatrix}, \quad\quad    \mathcal{C}_2 = -\begin{pmatrix}
        C_2\cdot D_2 & C_2 \cdot D_4 \\
        C_4\cdot D_2 & C_4 \cdot D_4 \\
    \end{pmatrix}.
\end{equation}
Computing the triple intersection numbers as in Appendix \ref{app: resolution} we obtain:
\begin{equation}\label{D4 cartan 1}
    \mathcal{C}_1 = \begin{pmatrix}
        4 & -2 \\
        -2 & 2 \\
    \end{pmatrix}, \quad\quad    \mathcal{C}_2 = \begin{pmatrix}
       4 & -2 \\
        -2 & 2 \\
    \end{pmatrix}. 
\end{equation}
Notice that the symmetrized Cartan matrices \eqref{D4 cartan 1} are precisely twice the symmetrized Cartan matrix of $B_2$. It is easy to check that on top of the origin the curve $C_1$ is a union of two distinct curves (and analogously for $C_2$), each with normal bundle $\mathcal{O}(0)\oplus\mathcal{O}(-2)$ (thus yielding the ``4'' entry in the Cartan matrices $\mathcal{C}_1$ and $\mathcal{C}_2$). Along the non-compact divisor $D_1$, but not on top of the origin, the curves recombine into a single irreducible curve. This is what leads to the folding from the naive simply-laced symmetry $D_3\cong A_3$ to the non-simply laced physical symmetry $B_2$. This mechanism is reminiscent of the monodromy action introduced by deformations of non-isolated singular lines in F-theory, resulting in non-simply laced symmetries, and studied e.g.\ in \cite{Grassi:2018wfy}. We plan on elucidating this relationship in further detail in future work.

The compact divisor $S_1$ can be characterized computing the square of its canonical bundle, finding:
\begin{equation}\label{KS1}
    K_{S_1}^2 = 2,
\end{equation}
implying that it is a degree 2 del Pezzo surface (namely $\mathbb{P}^2$ blown-up at 7 points, or $dP_7$).\\

\indent Summarizing, the resolved threefold $X_{\text{smooth}}$ features the homological data:
\begin{equation}
    \text{dim}\hspace{0.05cm}H_2(X_{\text{smooth}},\mathbb{R}) = 8, \quad\quad  \text{dim}\hspace{0.05cm}H_4(X_{\text{smooth}},\mathbb{R}) = 1,
\end{equation}
where the 8 independent curves come from the 7 blow-ups in the del Pezzo 2 surface, along with the hyperplane section. Translating the homological input into 5d SCFT data, we find:
\begin{equation}
    r = 1, \quad\quad f = 7,
\end{equation}
where the flavor symmetry manifest from the geometry is  $B_2\oplus B_2$, due to the singular non-compact lines.
We conclude that the theory engineered by the threefold \eqref{D4 example 1 def} has features compatible with the $E_7$ Seiberg theory. This would imply that the flavor symmetry of $X_{\text{smooth}}$ admits a non-abelian completion to $E_7$. The presence of a $B_2\oplus B_2$ subalgebra is in perfect agreement with the branching rules of $E_7$.\\
\indent Finally, notice that the threefold in \eqref{D4 example 1 def} does \textit{not} descend to a 4d class-$\mathcal{S}$ fixture. Indeed, a class-$\mathcal{S}$ setup of type $D_4$ on a sphere with three regular punctures, of which two of type $[5,1^3]$ (needed to engineer the flavor symmetry $B_2\oplus B_2$) is bad. Furthermore, the threefold contains a deformation term of the form $c_3z^2$, that cannot be reproduced by a class-$\mathcal{S}$ puncture, via the construction of Section \ref{sec: fixtures and class S}. Hence the circle reduction of the 5d SCFT at hand should be some other kind of 4d $\mathcal{N}=2$ SCFT, whose study we postpone to future work. We remark that these theories might be related to the 4d $\mathcal{N}=2$ theories recently studied in \cite{Bao:2026hjz}, arising from ``hidden Higgsings'' of class-$\mathcal{S}$ theories (namely, Higgsing of flavor symmetries that are not manifest from the punctures of the fixture).

\subsection{5d fixture that descends to a 4d fixture}\label{sec: 4d fixture}
In this Section we consider another example of CY3 of type \eqref{D4 example 1}: we are interested in turning on dynamical deformations, and thus producing a 5d fixture, in such a way that its dimensional reduction descends to a class-$\mathcal{S}$ setup of type $D_4$ with three regular punctures. We have shown in Section \ref{sec: fixtures and class S} that turning on deformations in \eqref{D4 example 1} is in one-to-one correspondence with partially closing the punctures. We turn on the dynamical deformation parameters in \eqref{def par D4} in such a way to obtain the following threefold:
\begin{equation}\label{D4 example 2 def}
\begin{cases}
    x^2+zy^2-z^3 + c_1uz^2=0,\\
    uv=x,
\end{cases}
\end{equation}
which should correspond to the class-$\mathcal{S}$ fixture:
\begin{equation}\label{class S def 2}
    ([7,1],[5,1^3],[3^2,1^2]).
\end{equation}
Coherently with the previous example, we now expect to break only the symmetry related to one of the non-compact singular lines. Indeed, it can be swiftly checked that the line $L_1: x=y=z=u=0$ supports an unscathed $D_4$ singularity, whereas the line $L_2: x=y=z=v=0$ sports a singularity of type $D_3\cong A_3$. As before, we will see that $L_2$ actually engineers a flavor symmetry algebra of type $B_2$. Performing an explicit complete crepant resolution, employing the methods reviewed in Appendix \ref{app: resolution}, produces 3 compact exceptional divisors, along with 6 non-compact ones. Their intersection pattern is depicted in Table \ref{table intersection}, where ``1'' means a non-trivial intersection, and ``0'' indicates that the two divisors do not intersect. We denote with $S_i, i=1,2,3$ the compact divisors, and with $D_i, i=1,\ldots,6$ the non-compact ones.

\begin{equation}\label{table intersection}
 \scalemath{0.9}{ \begin{array}{cccccccccc}
    &  S_1 & S_2 & S_3 & D_1 & D_2 & D_3 & D_4 & D_5 & D_6\\
    \hline
S_1 &     * & 1 & 1 & 1 & 0 & 0 & 1 & 1 & 1 \\
S_2 &     1 &*  & 1 & 1 & 1 & 1 & 0 & 1 & 1 \\
S_3 &     1 & 1 &*  & 1 & 1 & 1 & 1 & 1 & 1 \\
D_1 &     1 & 1 & 1 &*  & 1 & 1 & 1 & 0 & 0 \\
D_2 &     0 & 1 & 1 & 1 &  *& 0 & 0 & 0 & 0 \\
D_3 &     0 & 1 & 1 & 1 & 0 &*  & 0 & 0 & 0 \\
D_4 &     1 & 0 & 1 & 1 & 0 & 0 &*  & 0 & 0 \\
D_5 &     1 & 1 & 1 & 0 & 0 & 0 & 0 & * & 1 \\
D_6 &     1 & 1 & 1 & 0 & 0 & 0 & 0 & 1 &*  \\
\end{array}}
\end{equation}
Notice that the divisors $(D_1,D_2,D_3,D_4)$ manifestly intersect precisely like the $D_4$ Dynkin diagram. They originate from the resolution of the singular line $L_1$. On the other hand, $D_5$ and $D_6$ intersect compatibly with the $B_2$ Dynkin diagram, arising from the resolution of the singular line $L_2$. In the same fashion as in the example of Section \ref{sec: 5d fixture non simply laced}, the naive $D_3$ singularity supported by $L_2$ has undergone a folding, producing a $B_2$ flavor algebra. This can be rigorously proven computing the symmetrized Cartan matrix, exactly as shown in the previous section. The flavor curves are:
\begin{equation}
    C_5 = D_5\cdot (S_1+S_2), \quad\quad C_6 = D_6\cdot (S_1+S_2).
\end{equation}
Then the symmetrized Cartan matrix reads, following the procedure of Appendix \ref{app: resolution}:
\begin{equation}
    \mathcal{C} = -\begin{pmatrix}
        C_5\cdot D_5 & C_5 \cdot D_6 \\
        C_6\cdot D_5 & C_6 \cdot D_6 \\
    \end{pmatrix} =  \begin{pmatrix}
        4 & -2 \\
        -2 & 2 \\
    \end{pmatrix},
\end{equation}
which is precisely twice the symmetrized Cartan matrix of $B_2$.\\

In order to bridge the gap with the class-$\mathcal{S}$ description, we inspect the compact exceptional divisors in closer detail. Computing the square of the canonical bundle, as reviewed in Appendix \ref{app: resolution}, one gets:
\begin{equation}
    K^2_{S_1} = 5, \quad\quad  K^2_{S_2} = 4, \quad\quad  K^2_{S_3} = 6.
\end{equation}
This implies that $S_1,S_2,S_3$ are respectively a $dP_4,dP_5,dP_3$ surface. They respectively possess $5,6,4$ curve classes. These are not all independent, though, since the compact surfaces intersect: it can be checked that there are 4 curves lying at the intersection, which should not be counted twice. Hence only 11 curves are genuine independent elements of the second homology of the resolved threefold. All in all, we obtain the homological data of the resolved threefold $X_{\text{smooth}}$:
\begin{equation}
    \text{dim}\hspace{0.05cm}H_2(X_{\text{smooth}},\mathbb{R}) = 11, \quad\quad  \text{dim}\hspace{0.05cm}H_4(X_{\text{smooth}},\mathbb{R}) = 3,
\end{equation}
Therefore we find that the 5d SCFT engineered by \eqref{D4 example 2 def} has:
\begin{equation}\label{flavor rank 2}
    r = 3, \quad\quad f = 8.
\end{equation}
As we have shown, the flavor symmetry contains the subalgebra:
\begin{equation}\label{flavor subalgebra 2}
    G_{\text{flavor}} \supset D_4\oplus B_2.
\end{equation}
This result is in complete agreement with the data expected from the corresponding class-$\mathcal{S}$ setup, encoded by the 6d $\mathcal{N}=(2,0)$ theory of type $D_4$, reduced on a sphere with the three regular punctures in \eqref{class S def 2}.
The $[7,1]$ puncture engineers the $D_4$ symmetry, while the $[5,1^3]$ produces the $B_2$ flavor algebra. Employing the expressions in \cite{Chacaltana:2012zy}, one readily finds that the dimension of the Coulomb branch of the theory dictated by the puncture data in \eqref{class S def 2} is 3, and that its flavor symmetry is \cite{Chacaltana:2011ze}:
\begin{equation}
    Spin(10)\times Sp(2)\times U(1),
\end{equation}
which is perfectly compatible with the subalgebra \eqref{flavor subalgebra 2} (since the non-abelian completion of the extra $U(1)$ factor accounts for the enhancement from $D_4$ to $D_5$), and the total flavor rank in \eqref{flavor rank 2}.

\section{Gauging 5d fixtures}\label{sec: gauging}
5d SCFT fixtures, much in tune with their 4d fixture counterparts (when they exist), can be gauged together to produce novel 5d SCFTs.
As an illustrative example, take two 5d SCFT fixtures, engineered by two threefolds, say $X_1$ and $X_2$. Assume that both 5d fixtures contain the same simple flavor subalgebra $G$, arising from a singular non-compact line in their corresponding threefold: call $G_1$ and $G_2$ these two (identical) subalgebras, one for each fixture. Under suitable conditions, a diagonal subalgebra $G \subset G_1\oplus G_2$ can be gauged, producing a new 5d SCFT. From the geometric point of view, the new 5d SCFT is engineered by the threefold $X_3$, obtained by gluing (in a way that preserves the Calabi-Yau condition) the singular lines in $X_1$ and $X_2$ that engineer $G_1$ and $G_2$, thus forming a $\mathbb{P}^1$. $X_3$ is obtained by shrinking the volume of such $\mathbb{P}^1$ to zero, eliminating all scales and thus recovering the SCFT phase. This procedure has been introduced in \cite{DeMarco:2023irn}, to which we refer the reader for further details. Here, we are primarily interested in applying it to the case of 5d SCFT fixtures with non-simply laced flavor group, producing a new 5d SCFT with non-simply laced flavor group.

\subsection{Gauging the simply-laced flavor symmetry}

\indent Consider the example introduced in Section \ref{sec: 4d fixture}, which has a manifest flavor symmetry $D_4\oplus B_2$, coming from the non-compact singular lines in the threefold \eqref{D4 example 2 def}. We wish to take two copies of this 5d fixture, with flavor symmetries:
\begin{equation}\label{D4 subalgebras}
    G_1 \supset D_4\oplus B_2, \quad\quad G_2 \supset D_4\oplus B_2.
\end{equation}
We now gauge the diagonal subalgebra of the two $D_4$ subalgebras in \eqref{D4 subalgebras}: this should produce another 5d theory, with flavor symmetry at least $B_2\oplus B_2$. In order to concretely realize the gauging operation geometrically, consider the following threefolds, engineering the two initial 5d fixtures:
\begin{equation}
    X_1: \begin{cases}
        x^2+zy^2-z^3+(c_1+c_2v^2) u z^2 =0,\\
        uv = x.\\
    \end{cases} \quad\quad
        X_2: \begin{cases}
        x'^2+z'y'^2-z'^3+(c_2+c_1v'^2) u' z'^2 =0,\\
        u'v' = x'.\\
    \end{cases}
\end{equation}
Notice that we have added irrelevant terms in the defining equations of both threefolds ($c_2v^2uz^2$ and $c_1v'^2u'z'^2$, respectively), that leave the singularity and its resolution completely unaltered. The singular lines engineering the $D_4$ flavor symmetries in the two threefolds are:
\begin{equation}
 L_1:   x = y = z = u = 0, \quad\quad L_2: x'=y'=z'=u' = 0.
\end{equation}
Gauging the diagonal subgroup of the two $D_4$ flavor symmetries amounts to compactifying the two singular lines along a $\mathbb{P}^1$, which is realized by imposing the transition functions:
\begin{equation}\label{transition functions}
\begin{cases}
    x = x',\\
    y = y',\\
    z = z',\\
    v = \frac{1}{v'},\\
    u = v'^2 u'.\\
\end{cases}
\end{equation}
Clearly, $X_1$ and $X_2$ are identified under the transition functions \eqref{transition functions}. The SCFT phase is reached shrinking to zero the volume of the $\mathbb{P}^1$ that results from the identification of $L_1$ and $L_2$. This is implemented by the blow-down maps:
\begin{equation}\label{blowdown map}
    a = u', \quad b = v'^2 u', \quad x = u'v'.
\end{equation}
We hence obtain the final threefold $X_3$:
\begin{equation}\label{X3 threefold}
    X_3: \begin{cases}
        x^2+zy^2-z^3+c_1bz^2+c_2az^2 = 0,\\
        ab = x^2.\\
    \end{cases}
\end{equation}
Let's inspect the final result: notice that $X_3$ could have equivalently been obtained by gluing the undeformed versions of $X_1$ and $X_2$ (namely, setting $c_1 = c_2 = 0$), and \textit{then} turning on the deformation terms $c_1bz^2$ and $c_2az^2$. To see how this comes about, consider the undeformed threefolds:
\begin{equation}
    \hat{X}_1: \begin{cases}
        x^2+zy^2-z^3=0,\\
        uv = x.\\
    \end{cases} \quad\quad
        \hat{X}_2: \begin{cases}
        x'^2+z'y'^2-z'^3 =0,\\
        u'v' = x'.\\
    \end{cases}
\end{equation}
Imposing the transition functions \eqref{transition functions}, and subsequently the blow-down map \eqref{blowdown map} yields:
\begin{equation}
    \hat{X}_3: \begin{cases}
        x^2+zy^2-z^3= 0,\\
        ab = x^2.\\
    \end{cases}
\end{equation}
$\hat{X}_3$ is a 5d CM threefold of type \eqref{5d CM CY3}, that can be deformed according to \eqref{5d CM full deformations}. Finally, we can choose $\mathcal{F}(u,v,y,z)$ in such a way to turn on the deformation parameters $c_1bz^2$ and $c_2az^2$, precisely retrieving the deformed threefold \eqref{X3 threefold}. All in all, this argument can be summarized in the commutative diagram in Figure \ref{fig: commuting diagram}.

\begin{figure}[htbp]
  \centering
   \begin{tikzpicture}
    % Define nodes at exact positions (x, y)
    \node (A) at (0, 3) {$\hat{X}_1,\hat{X}_2$};
    \node (B) at (4, 3) {$X_1,X_2$};
    \node (C) at (0, 0) {$\hat{X}_3$};
    \node (D) at (4, 0) {$X_3$};

    % Arrows
    \draw[->] (A) -- node[above] {deform} (B);
    \draw[->] (A) -- node[left]  {glue} (C);
    \draw[->] (B) -- node[right] {glue} (D);
    \draw[->] (C) -- node[below] {deform} (D);
  \end{tikzpicture}
  \caption{Start from the undeformed threefolds $\hat{X}_1$ and $\hat{X}_2$. Deforming them to $X_1,X_2$, and then gluing $X_1$ and $X_2$ along non-compact singular lines produces $X_3$ (after shrinking the glued line to zero volume). Equivalently, $X_3$ is reached by gluing $\hat{X}_1$ and $\hat{X}_2$ along non-compact singular lines, shrinking the volume of such line to zero, thus obtaining $\hat{X}_3$, and finally deforming $\hat{X}_3$ to $X_3$.}
  \label{fig: commuting diagram}
\end{figure}

Physically, $X_3$ engineers a 5d SCFT with flavor symmetry at least $B_2\oplus B_2$, corresponding to the two singular lines that have not been affected by the gluing operation. Recall that both $X_1$ and $X_2$ descend in 4d to a class-$\mathcal{S}$ fixture. Therefore we can interpret the 5d gluing as a gauging between two 4d fixtures, corresponding to the dimensional reduction of $X_1$ and $X_2$: the outcome is a 4d class-$\mathcal{S}$ setup with 4 regular punctures:
\begin{equation}\label{4d puncture molecule}
    ([5,1^3],[5,1^3],[3^2,1^2],[3^2,1^2]),
\end{equation}
where the two $[7,1]$ punctures, appearing in the initial fixtures \eqref{class S def 2}, have been conformally gauged. The circle reduction of $X_3$ then produces the 4d theory encoded by the puncture data in \eqref{4d puncture molecule}. This can be further checked by explicitly computing the rank of the 4d and 5d Coulomb branch, which yields 10 in both cases. \\

\indent Naturally, this type of gauging can be applied in full generality to all types of 5d fixtures, as long as they possess one full puncture (if we perform a one-sided gauging) or two full punctures (if we wish to gauge two punctures, which is the standard bifundamental conformal matter case, explored in \cite{Bourget:2026ono}). The resulting gauged threefolds then read, in general:
\begin{equation}
    \begin{cases}
        P_{\mathfrak{g}}(x,y,z) + \mathcal{F}(u,v,y,z) = 0,\\
        uv=w(x,y,z),\\
    \end{cases} \quad \subset \mathbb{C}^5,
\end{equation}
where $w(x,y,z)$ admits a factorization dictated by the gauging, and the deformations $\mathcal{F}(u,v,y,z)$ correspond to (partial) closures of the two punctures that have not been gauged.

\subsection{Gauging the non-simply laced flavor symmetry}
It is also possible to produce new 5d SCFTs by gauging the non-simply laced flavor symmetries, employing the same technology reviewed above.\\
\indent Consider the threefold:
\begin{equation}\label{initial threefold}
    X:\quad \begin{cases}
        x^2+zy^2-z^3+c_1 u z^2 =0,\\
        uv = x.\\
    \end{cases}
\end{equation}
corresponding to a class-$\mathcal{S}$ fixture with puncture data of type $([7,1],[5,1^3],[3^2,1^2])$. We wish to gauge the singular non-compact line parametrized by "$u$", that corresponds to the partially closed puncture $[5,1^3]$, supporting a $B_2$ flavor symmetry factor. Impose the transition functions on the $\mathbb{P}^1$ resulting after the compactification of the singular line:
\begin{equation}\label{transition functions 2}
\begin{cases}
    x = x',\\
    y = y',\\
    z = u'z',\\
    u = \frac{1}{u'},\\
    v = u' v'.\\
\end{cases}
\end{equation}
Clearly, this operation preserves the Calabi-Yau condition, since the normal bundle of the $\mathbb{P}^1$ is $\mathcal{O}\oplus\mathcal{O}(-1)\oplus\mathcal{O}(-1)$ in the fourfold ambient space. The blowdown map, that shrinks the $\mathbb{P}^1$ to zero volume recovering the SCFT phase, is implemented via:
\begin{equation}
    a = u'v', \quad  b = u'z',\quad c = z',\quad d = v'.
\end{equation}
The singular threefold engineering the 5d SCFT after the gauging is (setting the parameter $c_1=1$):
\begin{equation}\label{glued fixture}
    \begin{cases}
        d^2+by^2-b^3+bc=0,\\
        a c = b d.\\
    \end{cases}
\end{equation}
It is singular at:
\begin{equation}
\begin{array}{lcl}
    a=b=c=d=0 &\rightarrow &  A_2 \text{ singularity}\\
   a=b=d=c+y^2=0 & \rightarrow &  A_1 \text{ singularity}\\
      b=c=d=y=0 & \rightarrow &  D_4 \text{ singularity}\\
    \end{array},
\end{equation}
where the last singular line is the initial $D_4$ singularity of the threefold \eqref{initial threefold}, while the first two have arose as a consequence of the gauging. Notice that the operation we have just performed does \textit{not} have a corresponding class-$\mathcal{S}$ description, since it involves the gauging of a $B_2$ flavor factor, which does not correspond to the maximal puncture\footnote{More general gaugings are allowed in the presence of irregular punctures, that we do not engineer in this work. See \cite{Chacaltana:2011ze} for additional details.}. Moreover, the threefold \eqref{glued fixture} is precisely of the form that engineers 5d fixtures defined in \eqref{5d CM full deformations}. We plan on systematically investigating these more general  gluings in future work.

\section{The Higgs Branch of 5d fixtures}\label{sec: higgs branch}
In this Section we sketch the reasoning to compute the UV dimension of the Higgs branch of 5d SCFT fixtures directly from the engineering threefold, furnishing a practical recipe. This confirms the agreement with the class-$\mathcal{S}$ description, when available. We defer a more comprehensive treatment to future work.\\

\indent The starting point are 5d CM threefolds, which are engineered by \eqref{5d CM CY3}. We have explained in Section \ref{sec: 5d fixtures} how to explicitly write down their deformations, which take the form \eqref{5d CM full deformations}, that we recall here:
\begin{equation}\label{5d CM def}
    \begin{cases}
        P_{\mathfrak{g}}(x,y,z) + \mathcal{F}(u,v,y,z)= 0,\\
        uv=w(x,y,z)+\mathcal{D}(x,y,z),\\
    \end{cases} \quad \subset \mathbb{C}^5.
\end{equation}
After turning on a subset of the deformations $\mathcal{F}(u,v,y,z)$, we construct a new threefold:
\begin{equation}\label{5d fixture def}
    \begin{cases}
        P_{\mathfrak{g}}(x,y,z) + \mathcal{F}_0(u,v,y,z) = 0,\\
        uv=w(x,y,z),\\
    \end{cases} \quad \subset \mathbb{C}^5,
\end{equation}
for some choice of $\mathcal{F}_0(y,z,u,v)$, which is a specialization of $\mathcal{F}(y,z,u,v)$. To compute the UV Higgs branch dimension of \eqref{5d fixture def}, we have to ask: how many dynamical deformations of \eqref{5d CM CY3} have been obstructed by turning on $\mathcal{F}_0(y,z,u,v)$?
Clearly, the dynamical deformations of the threefold \eqref{5d fixture def} take the shape:
\begin{equation}\label{5d fixture def 2}
    \begin{cases}
        P_{\mathfrak{g}}(x,y,z) + \mathcal{F}_0(u,v,y,z) +\mathcal{F}'(y,z,u,v)= 0,\\
        uv=w(x,y,z)+\mathcal{D}'(x,y,z)\\
    \end{cases} \quad \subset \mathbb{C}^5,
\end{equation}
for some $\mathcal{F}'(y,z,u,v)$ and $\mathcal{D}'(x,y,z)$. Since deforming the $u$ and $v$ singular lines by turning on $\mathcal{F}_0(y,z,u,v)$ does not affect the second equation, we expect that:
\begin{equation}
    \mathcal{D}'(x,y,z) = \mathcal{D}(x,y,z).
\end{equation}
It remains to be ascertained what is the precise form of $\mathcal{F}'(y,z,u,v)$, and in particular how many independent deformations it contains. We claim that, as long as the starting 5d fixture threefold \eqref{5d fixture def} engineers a 5d theory with non-trivial Coulomb branch rank:\\

\indent \textit{The dynamical deformations $\mathcal{F}'(y,z,u,v)$ are precisely those deformations in $\mathcal{F}(y,z,u,v)$ that decrease the rank of the 5d SCFT fixture engineered by the threefold \eqref{5d fixture def}}. \\

We can test this proposal explicitly. Take the 5d fixture engineered in Section \ref{sec: 4d fixture}, which has rank 3. It is defined by the threefold:
\begin{equation}\label{5d fixture HB}
\begin{cases}
    x^2+zy^2-z^3 + c_1uz^2=0,\\
    uv=x,
\end{cases}
\end{equation}
and hence $\mathcal{F}_0(y,z,u,v) =c_1uz^2$. It is easy to check, via an explicit resolution, that the only deformations in $\mathcal{F}(y,z,u,v)$ that \textit{do not} affect the rank of the 5d fixture are proportional to the monomials:
\begin{equation}
    (uz^2,u^4,u^5,u^2z,u^3z,u^3y).
\end{equation}
All the remaining deformations decrease the rank of the theory, and hence are still dynamical. There are 22 of them:
\begin{equation}
\begin{split}
& \mathcal{F}'(u,v,y,z) \ni \quad (z^2,vz^2,z,uz,vz,v^2z,v^3z,1,u,v,u^2,v^2,\\
 & \hspace{3.2cm}  u^3,v^3, v^4,v^5,y,uy,vy,u^2y,v^2y,v^3y). \\
 \end{split}
\end{equation}
Namely, the 4 deformations that do not depend on $u$ and $v$ are still dynamical, the 12 deformations depending on $v$ have been untouched, and only 6 out of the 12 deformations depending on $u$ are still dynamical. All in all, the expected dimension of the UV Higgs branch of the 5d fixture engineered by \eqref{5d fixture HB} is 25, also taking into account the 3 deformations contained in $\mathcal{D}(x,y,z)$. We can compare this result with the Higgs branch computed from its class-$\mathcal{S}$ reduction. We can employ the following expression \cite{Chacaltana:2012zy}, that yields the Higgs branch dimension in terms of the three regular punctures:
\begin{equation}\label{HB 4d}
    \operatorname{dim}_{\mathbb{H}} \mathrm{HB}\left(\mathcal{T}_{4 d}\right)=\frac{1}{2} \sum_i\left(\operatorname{dim}_{\mathbb{C}}(\mathfrak{g})-\operatorname{rank}(\mathfrak{g})-\operatorname{dim}_{\mathbb{C}}\left(\mathcal{O}_i^L\right)\right)+\operatorname{rank}(\mathfrak{g}),
\end{equation}
where $\mathcal{O}_i^L$ is the Spaltenstein dual of $\mathcal{O}_i$. Inputting the data of the 4d fixture \eqref{class S def 2}, it is easy to check that \eqref{HB 4d} yields a 25-dimensional Higgs branch, confirming the prediction from the 5d parent.

\section{Outlook}\label{sec: outlook}
In this note we have reduced M-theory on a class of canonical threefolds to construct 5d SCFT fixtures, which generalize the concept of 4d class-$\mathcal{S}$ fixtures with regular punctures. These novel 5d theories are interpreted as arising from the Higgs branch of bifundamental 5d conformal matter theories, where the bifundamental symmetry is produced by two full punctures in class-$\mathcal{S}$ (when such description is available). We have seen that not all 5d fixtures descend, upon dimensional reduction, to a 4d fixture. Indeed, we have provided a recipe to build singular threefolds corresponding to \textit{any} nilpotent orbit closure of the maximal punctures, irrespective of whether such operation produces a good, broken or bad 4d theory. Clearly, we wish to understand in greater detail what are the 4d descendants of 5d fixtures that do not admit a class-$\mathcal{S}$ fixture description.\\
\indent From the point of view of the classification of 5d SCFTs, we have at least two open tasks, in increasing degree of difficulty:
\begin{itemize}
    \item classify all 5d SCFT fixtures, organizing them into families labelled by their flavor algebras (with particular focus on the non-simply laced symmetries), following the work carried out for the 5d conformal matter cases, also relating them to known constructions of 5d SCFTs with non-simply laced flavor symmetry \cite{VanBeest:2020kxw,Bhardwaj:2019fzv,Tian:2021cif};
    \item explore the implications of the existence of 5d SCFT fixtures for the program of the classification of 5d SCFTs that admit a geometric origin. By construction, 5d fixtures come with an array of flavor symmetries that comprises all the simple Lie algebras. Analogously, 5d SCFTs with arbitrary rank can be produced via gauging of fixtures. It is then only natural to ponder whether they comprise all known geometric 5d SCFTs.
\end{itemize}
Finally, as we have mentioned, 5d fixtures come from Higgsings of bifundamental 5d conformal matter theories. For the 5d fixtures originating from bifundamentals of type $D$, and that admit a class-$\mathcal{S}$ reduction, an orthosymplectic magnetic quiver description is also available. We wish to investigate the quiver subtraction algorithms for such theories, mapping them to specific deformations, as done in the $A$-algebra case in \cite{DeMarco:2026jyj}.

\section*{Acknowledgments}
We would like to thank Jacques Distler and Riccardo Comi for insightful comments.
The work of AS and MDZ is supported by the VR Centre for Geometry and Physics
(VR grant No. 2022-06593). MDZ also acknowledges support from the Simons Foundation International (Simons Collaboration on Global Categorical Symmetries) and the VR project grant No. 2023-05590. This project has been partially funded by the European Union (ERC, HIGH, 101171852). Views and opinions expressed are however those of the author(s) only and do not necessarily reflect those of the European Union or the European Research Council. Neither the European Union nor the granting authority can be held responsible for them.

\appendix

\section{Crepant resolution in toric ambient space}
\label{app: resolution}
In this Appendix we briefly summarize the explicit technique employed to perform the complete crepant resolution of the canonical threefolds employed in the main text. The resolution works by blowing-up the singularities in a toric ambient space, in such a way that the Calabi-Yau condition is preserved. The intersection-theoretic computations described below have
been carried out with the aid of \textit{SageMath}, and the scripts are available from the authors upon request. For other works making use of this technique, we refer to \cite{Tian:2021cif,Mu:2023uws}. As an example, take the threefold in \eqref{D4 example 1 def}, that we report here for convenience:
\begin{equation}
\begin{cases}
    x^2+zy^2-z^3 + c_1uz^2+c_2vz^2+c_3z^2=0,\\
    uv=x,
\end{cases}
\end{equation}
which is equivalent to:
\begin{equation}\label{D4 example app}
 X:\quad    (uv)^2+zy^2-z^3 + c_1uz^2+c_2vz^2+c_3z^2=0.
\end{equation}
We blow-up the locus $(y,z,u,v)\rightarrow (y_1\delta_1,z_1\delta_1^2,u_1\delta_1,v_1\delta_1)$, embedding the coordinates in a toric ambient space, that imposes a $\mathbb{C}^*$-action on them with weights $\omega_i=(1,2,1,1)$. In order to guarantee that the resolution is crepant, the sum of the weights must satisfy: 
\begin{equation}
   \sum_i\omega_i = \omega_X +1, 
\end{equation}
where $\omega_X$ is the quasi-homogeneous weight under the $\mathbb{C}^*$-action of the polynomial defining $X$.
The blown-up threefold $X_1$ is a hypersurface in a toric ambient space:
\begin{equation}\label{toric action D4}
X_1: \quad \scalemath{0.85}{u_1^2 v_1^2+c_1 u_1 z_1^2 \delta_1+z_1 y_1^2+z_1^2 (c_3+c_2 v_1
   \delta_1+z_1 \delta_1^2)=0} \in \quad   \begin{array}{ccccc}
      y_1  & z_1  & u_1 & v_1 & \delta_1  \\
    \hline
    1 &  2 & 1  & 1 & -1 \\
    \end{array}
\end{equation}
Notice that $X_1$ is Calabi-Yau, since it is a weight $4$ hypersurface in a toric ambient space with canonical bundle $\mathcal{O}(-4)$, ensuring that the canonical bundle of $X_1$ is trivial by adjunction.
It is straightforward to check that the threefold $X_1$ is still singular. Iterating the blow-up procedure until no singularities are left produces the following smooth threefold, which is again given by a hypersurface in a toric ambient space:
\begin{equation}\label{X smooth}
\begin{array}{c}
 X_{\text{smooth}}: \quad \scalemath{0.85}{ U^2 V^2+ c_1 U Z^2 \delta_1\delta_2
   \delta_4^2+Z \left(\delta_2 \delta_3
   Y^2+c_3Z+ Z \delta_1\delta_3
 \left(c_2 V+ Z \delta_1\delta_2 \delta_4^2 \right)  \delta_5^2 \right) =0,} \\
   \rotatebox[origin=c]{270}{\ensuremath{\in}}\\
 \scalemath{0.9}{ \begin{array}{ccccccccc}
      Y & Z & U & V & \delta_1 & \delta_2 & \delta_3 & \delta_4 & \delta_5  \\
    \hline
    1 & 2 & 1 & 1 & -1 & 0 & 0 & 0 & 0 \\
    1 & 1 & 1 & 0 & 0 & -1 & 0 & 0 & 0 \\
    1 & 1 & 0 & 1 & 0 & 0 & -1 & 0 & 0 \\
    0 & 1 & 1 & 0 & 0 & 1 & 0 & -1 & 0 \\
    0 & 1 & 0 & 1 & 0 & 0 & 1 & 0 & -1 \\
    \end{array}}
\end{array}
\end{equation}
The toric ambient space comes with the following Stanley-Reisner ideal, which is computed keeping track of which loci are excluded at each blow-up step:
\begin{equation}\label{SR ideal}
\begin{split}
    I_{SR} = &\scalemath{0.85}{(\delta_2\delta_3,\delta_2\delta_5,\delta_2V,\delta_3\delta_4,\delta_3U,\delta_4\delta_5,\delta_4V,\delta_4Y,\delta_5U,\delta_5Y,}\\
    &\scalemath{0.85}{\delta_2\delta_5U,\delta_2UZ,\delta_3VZ,\delta_4VY,UYZ,VYZ)}.
    \end{split}
\end{equation}
To extract the data of the 5d SCFT engineered by $X_{\text{smooth}}$ we must examine its exceptional locus. There is one compact divisor given by $\delta_1=0$, along with four non-compact divisors, related to the vanishing of $\delta_2,\delta_3,\delta_4,\delta_5$. Their defining ideals are:
\begin{equation}
\begin{split}
   & S_1 = (\delta_1,U^2 V^2+c_3Z^2+Y^2 Z \delta_2 \delta_3),\\
&    D_1 = (\delta_2,U^2 V^2+c_3Z^2+V Z^2 \delta_1 \delta_3
   \delta_5^2),\\
      &  D_2 = (\delta_3,U^2 V^2+c_3Z^2+U Z^2 \delta_1
   \delta_2 \delta_4^2),\\
 &  D_3 = (\delta_4,U^2 V^2+c_3Z^2+Y^2 Z
   \delta_2 \delta_3+V Z^2 \delta_1 \delta_3
   \delta_5^2),\\
&   D_4 = (\delta_5,U^2 V^2+c_3Z^2+Y^2 Z \delta_2
   \delta_3+U Z^2 \delta_1 \delta_2 \delta_4^2).\\
   \end{split}
\end{equation}
Studying the intersection among divisors, one obtains the graph in Figure \ref{fig: D4 example 1}. Let's focus on the divisors $D_1$ and $D_3$ (the case of $D_2$ and $D_4$ is completely analogous). In order to detect the flavor symmetry supported by $D_1$ and $D_3$, we must compute the symmetrized Cartan matrix \eqref{sym cartan}, involving the curves defined in \eqref{D4 flavor curves 1}. The entries of the symmetrized Cartan matrix are triple intersection numbers between divisors \textit{in the threefold} $X_{\text{smooth}}$, which is not toric. It is more convenient, though, to pull-back the computation to the \textit{four-dimensional toric ambient space} $\mathcal{A}$ defined by the $\mathbb{C}^*$-action in \eqref{X smooth}: in this way, we can employ standard techniques applicable to toric varieties. With a slight abuse of notation, call $S_1,D_1,D_2,D_3,D_4$ the class related to the divisor (in the toric fourfold) given by $\delta_1=0,\delta_2=0,\delta_3=0,\delta_4=0,\delta_5=0$, respectively. Then the class of $X_{\text{smooth}}$ is:
\begin{equation}\label{class X smooth}
    [X_{\text{smooth}}] = -4S_1-2D_1-2D_2-4D_3-4D_4.
\end{equation}
The pulled-back $(1,1)$ entry of the Cartan matrix is:
\begin{equation}\label{quadruple intersection}
    \int_{X_{\text{smooth}}} C_1\cdot D_1 = \int_{\mathcal{A}}C_1\cdot D_1\cdot [X_{\text{smooth}}],
\end{equation}
and similarly for the other entries. We can now proceed to explicitly evaluating the quadruple intersection numbers in $\mathcal{A}$. Since $\mathcal{A}$ is non-compact, such intersection numbers are not well-defined, unless they involve at least one factor of the compact divisor $S_1$\footnote{We recall that a toric divisor $D_v$ is compact if and only if the star of the ray $v$ is a complete fan, equivalently if and only if $v$ lies in the
relative interior of the support of the fan of $\mathcal{A}$. In the case at hand only the ray associated with $\delta_1$ satisfies this criterion, the rays associated with $\delta_2,\dots,\delta_5$ lying on the boundary.}.
Hence we can reduce the computation of intersection numbers to the ambient space provided by the \textit{compact toric threefold} defined by the divisor $S_1$, with its fan $\Sigma_{S_1}$. To achieve this, we take the lattice $N_{\mathcal{A}}$ of the toric fourfold ambient space $\mathcal{A}$ and quotient it by the vector $v_{S_1}$ corresponding to $S_1$:
\begin{equation}\label{projection}
    N_{S_1} = N_{\mathcal{A}}/\mathbb{Z}v_{S_1}.
\end{equation}
The projection map $\phi$ is a homomorphism $\phi: \hspace{0.1cm} N_{\mathcal{A}} \rightarrow N_{S_1}$.
The fan $\Sigma_{S_1}$ is the star of the ray $v_{S_1}$: one takes the cones $\sigma \in \Sigma_{\mathcal{A}}$ having $v_{S_1}$ among their rays, and projects them to $N_{S_1} = N_{\mathcal{A}}/\mathbb{Z}v_{S_1}$ via $\phi$, as in \eqref{projection}. Since $S_1$ is compact, the resulting fan is complete. The rays of $\Sigma_{S_1}$ are the images $\phi(v_i)$ of those $v_i$ that share a cone with $v_{S_1}$. For all other divisors one has $\iota^{*}D_i = 0$, since
$D_i \cap S_1 = \emptyset$.

%The projection map $\phi$ is a homomorphism $\phi: \hspace{0.1cm} N_{\mathcal{A}} \rightarrow N_{S_1}$. The toric fan of $\Sigma_{S_1}$ is then constructed taking the cones in $\mathcal{A}$ that contain $S_1$ and projecting them via $\phi$. Since $S_1$ is a compact divisor in $\mathcal{A}$, it is guaranteed that the projected cones cover the whole of $\mathbb{R}^3$. 

The fourfold $\mathcal{A}$ is simplicial but not smooth: for instance the cone spanned by the rays associated with $\delta_1, Y, U, V$ has volume $2$ in the lattice, and correspondingly $\Sigma_{S_1}$ inherits a $\mathbb{Z}_2$ orbifold point. The hypersurface $X_{\rm smooth}$ nevertheless avoids these loci: setting $\delta_1 = Y = U = V = 0$ in \eqref{X smooth} leaves $c_3 Z^2$, which admits
no solution in the corresponding toric chart as long as $c_3 \neq 0$. Intersection numbers in $\Sigma_{S_1}$ are therefore rational in general, and the value attached to a maximal cone is the inverse of its volume. The numbers relevant for the 5d physics, obtained after contraction with $[X_{\rm smooth}]$, are integral.

%Furthermore, since $S_1$ is completely smooth in $\mathcal{A}$, so is the compact ambient space $\Sigma_{S_1}$. Degree-4 intersections in $\mathcal{A}$ are then pulled-back to degree-3 intersections in $\Sigma_{S_1}$. Here we can apply the standard toric notion concerning intersections: a triple intersection number is non-vanishing if and only if the three vectors defining it form a cone; its value is given by the inverse of the volume of the cone (so as to take account possible orbifold points). Then the triple intersection number is pushed-forward to the fourfold toric ambient space, multiplying by $S_1$. Intersection numbers involving powers of divisors greater than 1 are computed reducing them to triple intersections with powers at most 1, using the toric constraints given by relations among the columns of the ambient space $\mathcal{A}$. 

In the case at hand, we must compute e.g.\ \eqref{quadruple intersection}. Concretely, we proceed as follows. Let $\sigma$ be a degree-four monomial in the toric divisors of $\mathcal{A}$ containing at least one factor of $S_1$. By the projection formula:
\begin{equation}
\int_{\mathcal{A}} \sigma \;=\; \int_{\Sigma_{S_1}} \iota^{*}\!\left(\sigma/S_1\right),
\end{equation}
where $\iota : \Sigma_{S_1} \hookrightarrow \mathcal{A}$ and $\sigma/S_1$ denotes the monomial with one factor of $S_1$ removed. Any residual power of $S_1$ must be eliminated before restricting, since $S_1$ does not correspond to a ray of $\Sigma_{S_1}$: this is achieved using the linear equivalences among the columns of \eqref{X smooth}, which for the fan encoded by \eqref{X smooth} give in particular:
\begin{equation}
  S_1 \sim -(D_Y + D_1 + D_2 + D_3 + D_4).
\end{equation}

Inputting \eqref{class X smooth} and expanding the integral we rewrite \eqref{quadruple intersection} as:
\begin{equation}
    \int_{\mathcal{A}}C_1\cdot D_1\cdot [X_{\text{smooth}}]  = -2 D_1^3 S_1-2 D_1^2 D_2 S_1-4 D_1^2
   D_3 S_1-4 D_1^2 D_4 S_1-4 D_1^2
   S_1^2.
\end{equation}
The quadruple intersection numbers, computed first in the threefold $\Sigma_{S_1}$ as described above and then pushed-forward to the fourfold $\mathcal{A}$ are:
\begin{equation}
    D_1^3S_1 = 0,\quad  D_1^2D_2S_1 = 0,\quad D_1^2D_3S_1 =1, \quad D_1^2D_4S_1 = 0,\quad D_1^2S_1^2 = 0.
\end{equation}
Therefore we find:
\begin{equation}
    \int_{\mathcal{A}}C_1\cdot D_1\cdot [X_{\text{smooth}}] = -4.
\end{equation}
Repeating the computation for the other curves and non-compact divisors exactly reproduces the symmetrized Cartan matrices in \eqref{D4 cartan 1}. The same technology yields the square of the canonical bundle of the compact
divisors. Since $K_X = 0$, adjunction gives $K_{S_1} = S_1|_{S_1}$, hence
$K_{S_1}^2 = \int_{\mathcal{A}} S_1^{3}\cdot[X_{\rm smooth}]$, which is evaluated
exactly as above and reproduces \eqref{KS1}.

\section{Additional examples of 5d SCFT fixtures}\label{app: E8 example}
In this Appendix we include additional examples of 5d fixtures, obtained as a deformation of a 5d CM theory of type $E_6$ and $E_8$, respectively. We are particularly interested in exhibiting the deformations that Higgs the original flavor symmetry engineered by the non-compact lines. In particular, we show an example engineering a $G_2$ flavor symmetry.

\subsection{5d SCFT fixture of type $(E_6,SU(6))$}
As initial threefold, consider:
\begin{equation}\label{E6 atom}
    \begin{cases}
        x^2+y^3+\frac{z^4}{4}=0\\
        uv=z\\
    \end{cases},
\end{equation}
which engineers a 5d CM SCFT that descends to the 4d fixture encoded by the three regular punctures, specified by the Bala-Carter labels:
\begin{equation}
    (E_6,E_6,A_1).
\end{equation}
Consider now partially closing one of the maximal punctures, flowing to the 4d fixture:
\begin{equation}\label{E6 4d fixture}
    (E_6,E_6(a_1),A_1).
\end{equation}
The puncture $E_6(a_1)$ supplies a $SU(6)$ factor to the flavor symmetry \cite{Chacaltana:2014jba}. Applying the procedure detailed in Section \ref{sec: fixtures and class S} we can extract the threefold, which is a deformation of \eqref{E6 atom}, that engineers the parent 5d SCFT fixture. Considering $\Phi_u$ in a $A_1$ subalgebra of $E_6$ (namely, the Spaltenstein dual of $E_6(a_1)$) and computing the versal deformation parameters according to \eqref{spectral equations}, we find:
\begin{equation}\label{E6 fixture}
    \begin{cases}
        x^2+y^3+\frac{z^4}{4}-\frac{u}{2}yz^2+\frac{u^3}{24}z^2-\frac{u^4}{48}y+\frac{u^6}{864}=0,\\
        uv=z.\\
    \end{cases}
\end{equation}
The coefficients are crucial in order to enforce a singular line of type $SU(6)$. Indeed, the threefold \eqref{E6 fixture} is singular along:
\begin{equation}
   L_1: x=y=z=u=0, \quad\quad L_2: x=y-\frac{u^2}{12}=z=v=0,
\end{equation}
which respectively support a $E_6$ and $A_5$ singularity, as expected. Running the resolution method shown in Appendix \ref{app: resolution}, it is quickly checked that the threefold \eqref{E6 fixture} engineers a 5d SCFT with 4 compact divisors, coinciding with the rank of the 4d fixture encoded by the regular punctures in \eqref{E6 4d fixture}.

\subsection{5d SCFT fixture of type $(E_8,G_2)$}
\indent Consider the starting threefold:
\begin{equation}\label{E8 5d CM}
    \begin{cases}
        x^2+y^3+z^5 = 0, \\
        uv = z.\\
    \end{cases}
\end{equation}
As reviewed in Section \ref{sec: CM threefolds}, the threefold \eqref{E8 5d CM} engineers a 5d CM SCFT with flavor symmetry at least $E_8\oplus E_8$. Its circle reduction is a class-$\mathcal{S}$ fixture given by the 6d $\mathcal{N}=(2,0)$ theory of type $E_8$ on a Riemann sphere with three regular punctures, specified by the Bala-Carter labels:
\begin{equation}\label{E8 punctures}
    (E_8,E_8,A_1),
\end{equation}
where $E_8$ denotes the maximal puncture, and $A_1$ is the minimal puncture.
The allowed dynamical deformations of \eqref{E8 5d CM} are, according to Section \ref{sec: 5d fixtures}:
\begin{equation}\label{def par E8}
\begin{split}
 &  \mathcal{D}(x,y,z) \ni \quad(1), \\
  & \mathcal{F}(u,v,y,z) \ni \quad (yz^3,uyz^3,vyz^3,yz^2,uyz^2,vyz^2,u^2yz^2,v^2yz^2,u^3yz^2,v^3yz^2,u^4yz^2,v^4yz^2,\\
&u^5yz^2,v^5yz^2,u^6yz^2,v^6yz^2,u^7yz^2,v^7yz^2,z^3,uz^3,vz^3,u^2z^3,v^2z^3,u^3z^3,v^3z^3,u^4z^3,v^4z^3,u^5z^3,v^5z^3,\\
&u^6z^3,v^6z^3,u^7z^3,v^7z^3,u^8z^3,v^8z^3,u^9z^3,v^9z^3,u^{10}z^3,v^{10}z^3,u^{11}z^3,v^{11}z^3,yz,uyz,vyz,u^2yz,v^2yz,\\
&u^3yz,v^3yz,u^4yz,v^4yz,u^5yz,v^5yz,u^6yz,v^6yz,u^7yz,v^7yz,u^8yz,v^8yz,u^9yz,v^9yz,u^{10}yz,v^{10}yz,\\
&u^{11}yz,v^{11}yz,u^{12}yz,v^{12}yz,u^{13}yz,v^{13}yz,z^2,uz^2,vz^2,u^2z^2,v^2z^2,u^3z^2,v^3z^2,u^4z^2,v^4z^2,u^5z^2,v^5z^2,\\
&u^6z^2,v^6z^2,u^7z^2,v^7z^2,u^8z^2,v^8z^2,u^9z^2,v^9z^2,u^{10}z^2,v^{10}z^2,u^{11}z^2,v^{11}z^2,u^{12}z^2,v^{12}z^2,u^{13}z^2,v^{13}z^2,\\
&u^{14}z^2,v^{14}z^2,u^{15}z^2,v^{15}z^2,u^{16}z^2,v^{16}z^2,u^{17}z^2,v^{17}z^2,y,uy,vy,u^2y,v^2y,u^3y,v^3y,u^4y,v^4y,u^5y,v^5y,\\
&u^6y,v^6y,u^7y,v^7y,u^8y,v^8y,u^9y,v^9y,u^{10}y,v^{10}y,u^{11}y,v^{11}y,u^{12}y,v^{12}y,u^{13}y,v^{13}y,u^{14}y,v^{14}y,\\
&u^{15}y,v^{15}y,u^{16}y,v^{16}y,u^{17}y,v^{17}y,u^{18}y,v^{18}y,u^{19}y,v^{19}y,z,uz,vz,u^2z,v^2z,u^3z,v^3z,u^4z,v^4z,\\
&u^5z,v^5z,u^6z,v^6z,u^7z,v^7z,u^8z,v^8z,u^9z,v^9z,u^{10}z,v^{10}z,u^{11}z,v^{11}z,u^{12}z,v^{12}z,u^{13}z,v^{13}z,u^{14}z,v^{14}z,\\
&u^{15}z,v^{15}z,u^{16}z,v^{16}z,u^{17}z,v^{17}z,u^{18}z,v^{18}z,u^{19}z,v^{19}z,u^{20}z,v^{20}z,u^{21}z,v^{21}z,u^{22}z,v^{22}z,u^{23}z,v^{23}z,\\
&1,u,v,u^2,v^2,u^3,v^3,u^4,v^4,u^5,v^5,u^6,v^6,u^7,v^7,u^8,v^8,u^9,v^9,u^{10},v^{10},u^{11},v^{11},u^{12},v^{12},u^{13},v^{13},\\
&u^{14},v^{14},u^{15},v^{15},u^{16},v^{16},u^{17},v^{17},u^{18},v^{18},u^{19},v^{19},u^{20},v^{20},u^{21},v^{21},u^{22},v^{22},u^{23},v^{23},u^{24},v^{24},u^{25},v^{25},\\
&u^{26},v^{26},u^{27},v^{27},u^{28},v^{28},u^{29},v^{29}).\\
\end{split}
\end{equation}
$\mathcal{D}(x,y,z)$ contains 1 deformation parameter, while $\mathcal{F}(u,v,y,z)$ contains 248 independent deformation parameters.\\
\indent Pick the following deformed threefold:
\begin{equation}\label{E8 5d CM def}
      \begin{cases}
        x^2+y^3+z^5+c_1u^2z^3 = 0, \\
        uv = z.\\
    \end{cases}  
\end{equation}
Notice that \eqref{E8 5d CM def} is singular along two lines:
\begin{equation}\label{singular lines E8}
    L_1: x=y=z=u=0, \quad\quad L_2: x=y=z=v=0.
\end{equation}
The fiber on a generic point of $L_1$ is a $E_8$ singularity, as in \eqref{E8 5d CM}. On the other hand, $L_2$ supports a singularity of $D_4$ type. Employing the techniques outlined in Appendix \ref{app: resolution}, it can be shown that the complete crepant resolution of \eqref{E8 5d CM def} inflates 16 exceptional divisors: 6 are compact, and 10 are non-compact. Therefore the 5d SCFT engineered by \eqref{E8 5d CM def} is a rank 6 theory. The intersection pattern and the symmetrized Cartan matrix given by the flavor curves implies that the flavor symmetry manifest from the non-compact singular lines is:
\begin{equation}
    G_{\text{flavor}} \supset E_8 \oplus G_2.
\end{equation}
Physically, the deformation ``$+c_1u^2z^3$'' in \eqref{E8 5d CM def} has broken one of the initial $E_8$ symmetries to $G_2$. Geometrically, the $G_2$ symmetry comes from the folding of the exceptional divisors related to the $D_4$ singularity supported on $L_2$ in \eqref{singular lines E8}. The mechanism for producing non-simply laced symmetries is analogous to what was shown in Section \ref{sec: 5d fixture non simply laced} for the folding $A_3 \rightarrow B_2$.\\
\indent Correspondingly, the circle reduction of the 5d SCFT engineered by \eqref{E8 5d CM def} is the class-$\mathcal{S}$ theory of type \eqref{E8 punctures}, where one of the punctures has been partially closed, in order to produce a $G_2$ flavor symmetry:
\begin{equation}\label{E8 punctures 2}
     (E_8,D_6(a_1),A_1). 
\end{equation}
It is straightforward to check that the fixture \eqref{E8 punctures 2} is a rank 6 theory, coherently with the 5d picture.

\section{Review of cDV threefolds}\label{app: cDV}
In this Appendix we briefly review the machinery developed in \cite{Collinucci:2021ofd,DeMarco:2022dgh}, to which we refer the reader for the full details. Consider a special instance of cDV threefold, in which only parameters depending on the versal deformation of a Du Val singularity appear. This is defined as:
\begin{equation}\label{cDV appendix}
    P_{\mathfrak{g}}(x,y,z)+u\mathcal{F}(u,y,z) = 0,
\end{equation}
where the part of $\mathcal{F}(u,y,z)$ that depends on $(y,z)$ can only contain monomials that appear in the versal deformation \eqref{Du Val versal} of $\mathfrak{g}$. Given an algebra $\mathfrak{g}$ and a nilpotent orbit $\mathcal{O}_u$, we can construct a Higgs field $\Phi_u \in \mathfrak{g}$ as in \eqref{higgs field}, employing the Slodowy slice through a nilpotent element accompanying $\mathcal{O}_u$. The element $\Phi_u$ encodes the threefold equation \eqref{cDV appendix} via its Casimirs. More formally, the threefolds are defined as follows:
\begin{equation}\label{spectral equations}
\scalemath{0.9}{
    \begin{array}{ll}
       A_n: & \quad xy-\text{det}(z\mathbbm{1}+\Phi_u)=0\\
        D_n: &\quad x^2+zy^2-\frac{\sqrt{\text{det}(z\mathbbm{1}+\Phi_u^2)}-\text{Pfaff}^2(\Phi_u)}{z}+2y\hspace{0.1cm}\text{Pfaff}(\Phi_u)=0,\\
        E_{6}:& \quad x^{2}+y^{3}+\frac{z^4}{4}+\epsilon_{2}(u) y z^{2}+\epsilon_{5}(u) y z+\epsilon_{6}(u) z^{2}+\epsilon_{8}(u) y+\epsilon_{9}(u) z+\epsilon_{12}(u) =0,\\
       E_{7}:& \quad x^{2}+y^{3}+y z^{3}+\tilde{\epsilon}_{2}(u) y^{2} z+\tilde{\epsilon}_{6}(u) y^{2}+\tilde{\epsilon}_{8}(u) y z+\tilde{\epsilon}_{10}(u) z^{2}+\tilde{\epsilon}_{12}(u) y+\tilde{\epsilon}_{14}(u) z+\tilde{\epsilon}_{18}(u)=0, \\
E_{8}: &\quad x^{2}+y^{3}+z^{5}+\hat{\epsilon}_{2}(u) y z^{3}+\hat{\epsilon}_{8}(u) y z^{2}+\hat{\epsilon_{12}}(u) z^{3}+\hat{\epsilon}_{14}(u) y z+\hat{\epsilon}_{18}(u) z^{2}+\hat{\epsilon}_{20}(u) y+\hat{\epsilon}_{24}(u) z+\hat{\epsilon}_{30}(u)=0.\\
    \end{array}}
\end{equation}
The dependence of the parameters $\epsilon_i(u),\tilde{\epsilon}_i(u)$ and $\hat{\epsilon}_i(u)$ on the Casimirs of $\Phi_u$ is complicated, and it is spelled out explicitly in Appendix C of \cite{DeMarco:2022dgh}.

\clearpage

\bibliographystyle{at}
\bibliography{bibliography.bib}

\end{document}